# Dual antagonistic role of motor proteins in fluidizing active networks

Bibi Najma[1†], Minu Varghese[1,2†], Lev Tsidilkovski[1], Linnea Lemma[1,3,4], Aparna Baskaran[1], and Guillaume Duclos[1*]

[1]Department of Physics, Brandeis University, Waltham, MA 02453

[2]Department of Physics, University of Michigan, Ann Arbor, MI 48109

[3]Department of Physics, University of California at Santa Barbara, Santa Barbara, CA 93106

[4]Current address: Department of Chemical and Biological Engineering, Princeton University, Princeton, NJ 08544

[†] These authors contributed equally

*gduclos@brandeis.edu

**Abstract**
Cells accomplish diverse functions using the same molecular building blocks, from setting up cytoplasmic flows to generating mechanical forces. In particular, transitions between these non-equilibrium states are triggered by regulating the expression and activity of cytoskeletal proteins. However, how these proteins set the large-scale mechanics of the cytoskeleton and drive such non-equilibrium phase transitions remain poorly understood. Here, we show that a minimal network of biopolymers, molecular motors, and crosslinkers exhibits two distinct emergent behaviors depending on its composition, spontaneously flowing like an active fluid, or buckling like an active solid. Molecular motors play a dual antagonistic role, fluidizing or stiffening the network depending on the ATP concentration. By combining experiments, continuum theory, and chemical kinetics, we identify how to assemble materials with targeted activity and elasticity by setting the concentrations of each component. Active and elastic stresses can be further manipulated *in situ* by light-induced pulses of motor activity, controlling the solid-to-fluid transition. These results highlight how cytoskeletal stresses regulate the self-organization of living matter and set the foundations for the rational design and control of active materials.

**Main**
Active matter describes out-of-equilibrium materials composed of motile building blocks that convert free energy into mechanical work (*1*). The continuous input of energy at the particle scale liberates these systems from the constraints of thermodynamic equilibrium, leading to hydrodynamic instabilities not found in passive materials (*2-7*). Living cells are prototypical examples of adaptive multifunctional active materials (*8*). They self-organize in diverse out-of-equilibrium states using the same molecular machinery, from spontaneously flowing vortexes during cytoplasmic streaming (*9*) to polar asters during chromosome segregation (*10*). Similar self-organization has been reported in biomimetic active matter (*11-13*) and modeled using symmetry-based hydrodynamic theories (*1, 14, 15*). However, contrary to reconstituted systems, cells can spontaneously and reversibly trigger transitions between these non-equilibrium states by regulating the expression of their molecular building blocks. Understanding these non-equilibrium dynamical transitions at the molecular level is challenging. In particular, little is known about how to relate the forces generated at the microscopic scale to macroscopic emergent dynamics - both numerically and experimentally (*16-22*). As a result, the predictive assembly of active fluids with controllable material properties remains unexplored, despite recent advances in machine-learning driven forecasting of active nematics (*23, 24*), light-harvesting molecular motors (*25-27*), and the spatiotemporal control of active dynamics (*28-30*).

In this article, we show that investigating spontaneous deformations in a minimal *in vitro* system composed of cytoskeletal proteins can reveal how to rationally design and control active biomimetic materials. We combine experiments and theory to explore the origin of activity-driven instabilities in thin extensile networks composed of microtubule bundles, kinesin molecular motors, and crosslinkers. We show that depending on the molecular composition of the network, the extensile nature of the kinesin-microtubule bundles can drive two distinct instabilities: a bend instability, where the network spontaneously deforms in-plane, and a buckling instability, where the active network spontaneously buckles out-of-plane.

A hydrodynamical model of a thin active elastomer shows that the balance between activity and elasticity controls the directionality of the most unstable mode: active fluids spontaneously bend in-plane (*31*) while active solids buckle out-of-plane (*32, 33*). The transition from an active fluid to an active solid is set by the ratio of active motors to passive crosslinkers. We further estimated how the activity and the elasticity of the network are set by the concentrations of molecular motors and crosslinkers using an enzyme kinetics model. We show that motor proteins have a dual antagonistic role depending on ATP concentration: ATP-powered motor stepping fluidizes the crosslinked network (*34*), but between two power strokes, motors can also act as crosslinkers, enabling the transmission of elastic stresses (*35*). The multiscale mapping between the cytoskeletal composition and the mechanical properties of the material provides a quantitative estimate of the active and elastic stresses. Finally, we show how the mechanical properties can be controlled *in situ* using light-dimerizable molecular motors. Taken together, these results serve as a paradigm for the rational design and control of active matter, a requirement for any adaptive and reconfigurable materials applications.

**Thin extensile active networks spontaneously deform in-plane or out-of-plane depending on their molecular composition.**

While ordered fluids at thermodynamic equilibrium tend to minimize energy by uniformly aligning their elongated units, active nematics are intrinsically unstable (*36*). The energy injected at the particle scale drives the spontaneous growth of long-range deformations (*31*) and subsequent nucleation of topological defects (*37, 38*). Here, we assemble an active material composed of rod-like microtubules and kinesin molecular motors (**Fig. 1a**). Stabilized microtubules are either bundled by a depletant or crosslinked by PRC1, a microtubule-specific crosslinker (*39, 40*). The polymer network is driven away from equilibrium by the adenosine 5’ -triphosphate (ATP) fueled stepping of clusters of molecular motors. The motor clusters slide apart antiparallel microtubules, exerting dipolar extensile stresses that drive the spontaneous growth of deformations, leading to chaotic dynamics (*2, 41*) and active turbulence (*42, 43*).

**Fig.1: Spontaneous growth of two instabilities in 3D active networks of crosslinked microtubules.**

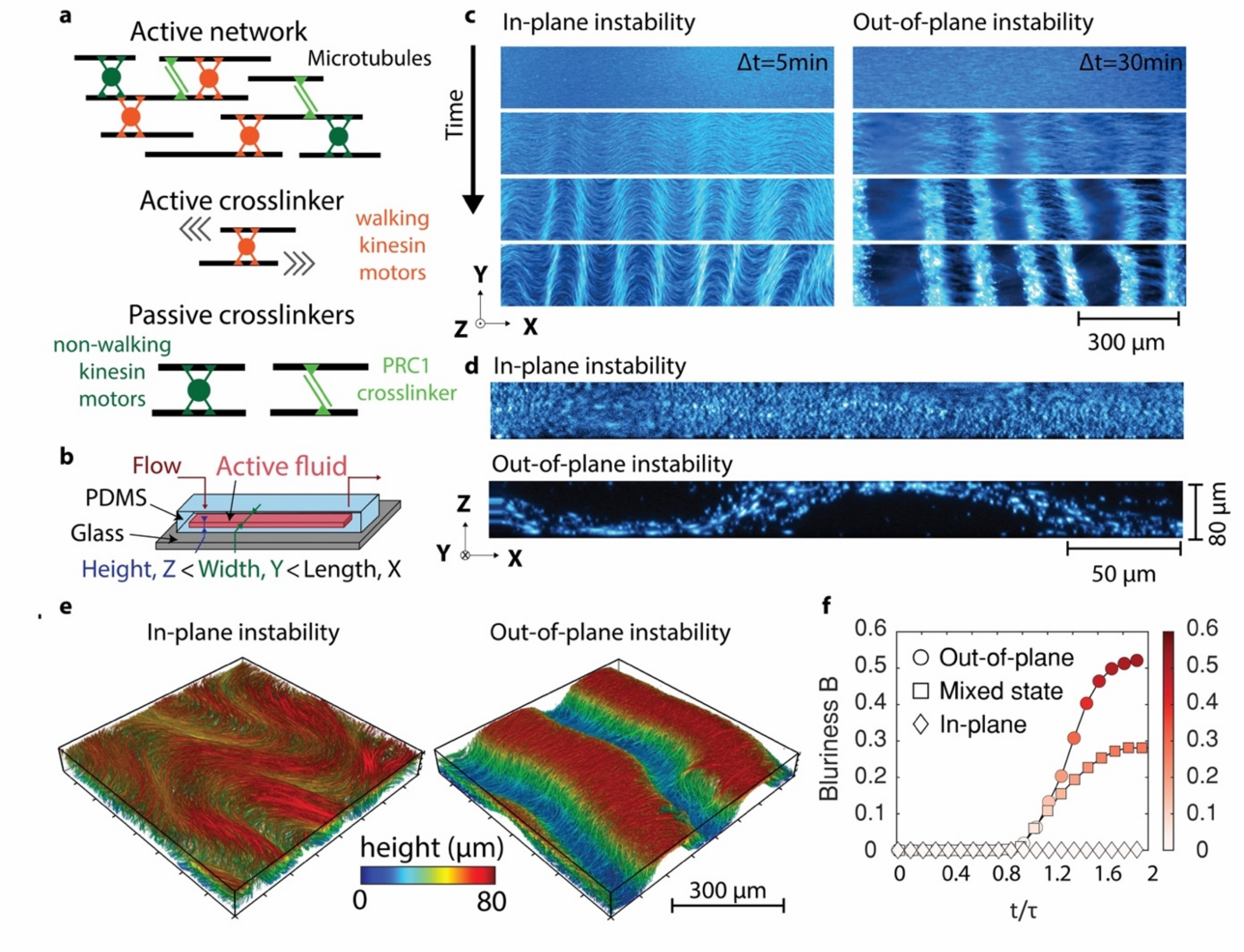

a) Schematic of an active network: microtubules are crosslinked by clusters of kinesin molecular motors and passive PRC1 crosslinker. In the presence of ATP, they form a stress–generating extensile bundle. In the absence of ATP, the molecular motors are not walking. Instead, they passively crosslink adjacent microtubules. b) Schematic of the microfluidic channels with thin rectangular cross-sections (H=80 µm, W=3 mm, L=3cm). c) Time series for the in-plane and out-of-plane instabilities starting from flow-aligned networks ([motor cluster]=20nM and respectively [ATP]=1.4 mM and 2µM, fluorescent widefield microscopy). d) Confocal cross-section along the direction of initial alignment of actively bending and buckling networks. e) 3D volume rendering of confocal images for the in-plane and out-of-plane instabilities. The color code shows optical sections taken at different depths. f) Time evolution of the blurriness coefficient B measured for three unstable networks. τ is the characteristic time for the growth of instability. The blurriness quantifies the fraction of the widefield fluorescent image that is out-of-focus. When the blurriness is null, the entire image is in focus and the instability is purely in-plane. The larger the blurriness, the more out-of-plane buckling is detected. The color map indicates the blurriness B.

We confined this active material within a thin microfabricated channel (**Fig. 1b**). We initially flowed the material within the channel to uniformly align the polymer bundles along the channel length, similarly to passive rod-like colloids under shear flows (*44, 45*). Then, we stopped the shear flow, at which point the aligned active network was unstable, and periodic deformations with a finite wavelength spontaneously grew perpendicular to the initial alignment (**Fig. 1c**). At high ATP concentrations, we observed the growth of in-plane deformations that result from the previously reported generic bend instability in extensile active liquid crystals (**Fig. 1c, Videos S1-S2, S5**) (*46*). Decreasing the concentration of ATP led to an unexpected transition: at low ATP concentrations, in-plane deformations are suppressed (**Fig. S1**), and regularly spaced patches of the network slowly grow out-of-focus (**Fig. 1c, Videos S3-S4)**. Confocal microscopy confirmed that the out-of-focus domains correspond to microtubule bundles buckling out-of-plane (**Fig. 1d-e**, **Video S6**). This instability is reminiscent of the out-of-plane buckling reported for suspensions of longer microtubules and lower motor concentrations (*32, 33*). Eventually, the active network evolved into a chaotic flow regime, irrespective of the directionality of the first instability.

To further characterize the directionality of the instability, we defined the blurriness coefficient B, a scalar varying between 0 and 1 that quantifies the relative area of the field-of-view that is out-of-focus (**Fig. S2**, **Video S7,** see SI). Briefly, if the instability is purely in-plane, then the whole image is in-focus and the blurriness is null (B=0, diamond symbols on **Fig. 1f**). If the instability is purely out-of-plane, a large portion of the image is out-of-focus (B>0.5, circle symbols on **Fig. 1f**) For intermediate values of blurriness, the instability is a superposition of in-plane and out-of-plane deformations (square symbols on **Fig. 1f**).

We hypothesize that the ATP-controlled transition from out-of-plane buckling to in-plane bending is a signature of a solid-to-fluid transition. Multiple clues motivate this hypothesis. First, active hydrodynamic theory has shown that the in-plane bend instability is a generic feature of 3D active nematic liquid crystal confined in thin channels (*4, 46*). Second, motor clusters will also bind to microtubules in the absence of ATP (*47*). Decreasing the concentration of ATP leads to motors dwelling longer on microtubules between two ATP-fueled power strokes, resulting in a larger proportion of motors passively crosslinking microtubules instead of sliding them apart. Third, recent rheology experiments showed that the elastic and loss moduli of isotropic kinesin-microtubule networks depend on the concentration of ATP (*35*); Finally, the out-of-plane buckling is reminiscent of the compressive winkling of thin elastic sheets (*48*).

**The balance between activity and crosslinking controls the directionality of the most unstable mode.**

To explore this hypothesis, we further investigated how the molecular composition of the active network controls the direction of the instability. Slowly increasing the ATP concentration triggered a sharp transition from an out-of-plane to an in-plane instability around 5µM (**Fig. 2a**). Measurements on passive visco-elastic gels composed of biopolymers suggest that increasing the number of crosslinkers should lead to stiffer networks (*49*). In agreement with this expectation, we observed that sparsely crosslinked networks spontaneously bent in plane while densely crosslinked networks buckled out-of-plane (**Fig. 2b**). When ATP was abundant, increasing the number of motor clusters induced a transition from an out-of-plane buckling to an in-plane

**Fig. 2: The transition from an in-plane to out-of-plane instability is set by the ratio of active motors to passive crosslinkers**

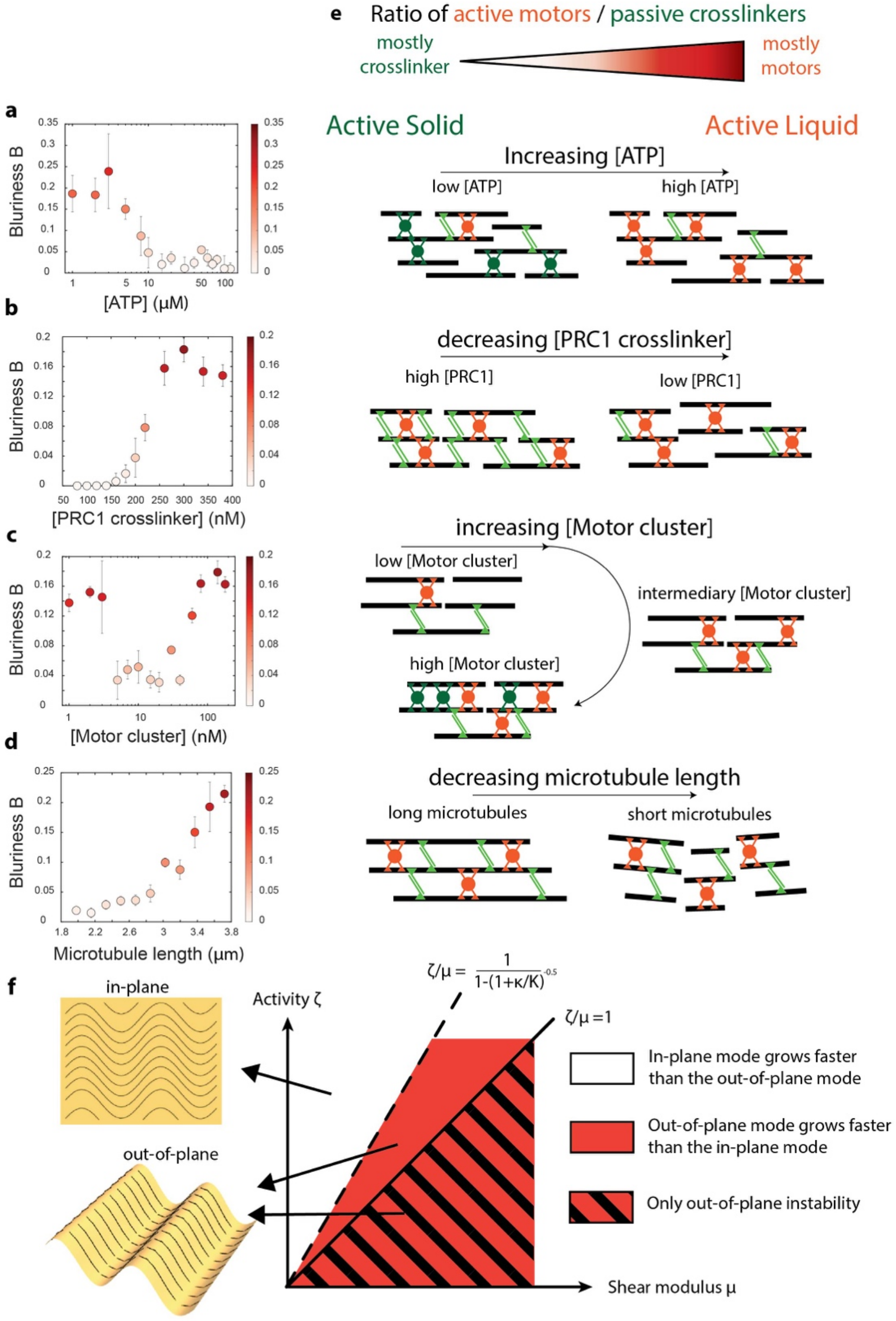

Blurriness B as a function of a) ATP concentration: increasing [ATP] leads to a transition from an out-of-plane (B>0) to an in-plane (B=0) instability ([motor cluster]=3nM, [PRC1 crosslinker]=100nM); b) PRC1 crosslinker concentration: increasing [PRC1 crosslinker] leads to a transition from an in-plane to an out-of-plane instability ([ATP]=40μM, [motor cluster]=60nM). c) motor clusters concentration: increasing [motor clusters] can induce a reentrant transition when [ATP] is low ([ATP]=15μM, [PRC1 crosslinker]=100nM). d) average microtubules' length ([ATP]=8μM, [motor cluster]=15nM, [PRC1 crosslinker]= 100nM). In a-d, the color map indicates the blurriness B, and the error bars represent the standard deviation over N>3 independent replicates; e) Schematics of an active network where the ratio of active motors to passive crosslinkers or the polymers' length are changed. Orange motor clusters are ATP-bound (aka active), while green motor clusters are passively crosslinking microtubules. f) Theoretical phase diagram showing how the interplay between activity $\zeta$ and shear modulus $\mu$ of a thin active elastomer sheet sets the direction of the most unstable mode. The sheet spontaneously deforms in 3D along the direction of the most unstable mode. In-plane deformations spontaneously grow only for $\zeta/_{\mu} > 1$ (continuous line) while the out-of-plane modes are always unstable.

instability (**Fig. 2c**). Interestingly, in the ATP limiting regime, increasing the number of motor clusters triggered a re-entrant transition from in to out-of-plane deformations (**Fig. 2c**). Finally, increasing the length of the microtubules while maintaining constant number of tubulin monomers led to a transition from in-plane instability to an out-of-plane bucking (**Fig. 2d**). **Fig. 2e** summarizes these observations, suggesting that the ratio of active motors (ATP-bound kinesin) to passive crosslinkers (PRC1 and non-ATP-bound kinesin) controls the directionality of the instability. Of note, we observed similar phase behaviors for active microtubule networks bundled by a depleting agent instead of a crosslinker (**Fig. S3**) and for networks powered by non-processive K-365 kinesin motor clusters that still slide bundles apart but detach from the microtubules after each step (**Fig. S4**).

**Hydrodynamic model for an active elastomer.**

To better understand how the interplay between activity and elasticity sets the direction of the instability, we modeled the active network as a thin sheet of active nematic elastomer in a quasistatic medium (*1, 32, 50*). We assumed that the sheet initially lays flat in the xy plane with nematic order along x̂, and considered small deformations of the form $\vec{u} = (u_x, u_y, h)$. Further, we assumed that $\vec{u}$ varies spatially only along x̂. Since the material points of the elastic sheet are the nematogens, the fluctuations in the director are coupled to the displacement field in the sheet. As a result, the nematic director in the deformed state can be written as $\vec{n} = \hat{x} + \partial_x \vec{u}$. The flat configuration of the sheet is unstable because a deformation $\vec{u}(\vec{r})$ at a point $\vec{r}$ on the sheet experiences destabilizing active forces $\zeta \vec{\nabla} \cdot (\vec{n}\,\vec{n}) = \zeta(\partial_x^2 u_y \hat{y} + \partial_x^2 h \hat{z})$ from its surroundings (*31*). The total free energy associated with the deformations is:

$$\mathcal{F}[u_x, u_y, h] = \frac{1}{2} \int dxdy \left[ \nu(\partial_x u_x)^2 + \mu\left(\partial_x u_y\right)^2 + \kappa(\partial_x^2 h)^2 + K \left[ \left(\partial_x^2 u_y\right)^2 + (\partial_x^2 h)^2 \right] \right] \quad (1)$$

where $\nu$ is the bulk modulus, $\mu$ is the shear modulus, $\kappa$ is the bending modulus, and K is the nematic elasticity (see SI). Therefore, the deformation experiences a restoring force $-\frac{\delta \mathcal{F}}{\delta \vec{u}(\vec{r})}$ from the rest of the elastic sheet. It also experiences an additional restoring force $-\gamma \partial_t \vec{u}(\vec{r})$ through friction internal to the material and from the surrounding fluid. Assuming the system is overdamped, the balance of active, elastic and frictional forces experienced by a material point gives the following dynamics for the in-plane $u_y$ and out-of-plane $h$ deformations:

$$\partial_t u_y = \frac{1}{\gamma}\,[(\mu - \zeta)\partial_x^2 - K\,\partial_x^4]\,u_y \quad (2)$$

$$\partial_t h = -\frac{1}{\gamma}\,[\zeta\partial_x^2 + (K + \kappa)\,\partial_x^4]h \quad (3)$$

Note that both equations are unstable when the activity $\zeta$ is large. Eq. 2 in the absence of elasticity is $\partial_t \vec{u} = -\frac{1}{\gamma}[\zeta\nabla \cdot \vec{n}\,\vec{n} + \nabla \cdot \sigma^p]$ where $\sigma^p$ is the passive stress arising from the nematic. Thus, it corresponds to the active flow that will be induced in the presence of strong substrate friction in the well-studied active nematic theories (*1, 50, 51*). This leads to an in-plane instability known as the generic instability in active nematics (*31*), a fluid-like instability. Eq. 3 precisely corresponds to height fluctuations in an elastic thin film and leads to an activity-driven out-of-plane buckling instability, reminiscent of Euler buckling in solids (*32, 33*). Eq. 3 shows that out-of-plane modes are always unstable at any non-zero activity. Eq. 2 predicts that in-plane modes are unstable only when the activity is larger than a critical activity $\zeta^* = \mu$, where $\mu$ is the shear modulus. The existence of a non-zero critical activity is a consequence of the elastic response of the active sheet, which is fundamentally different from critical activity resulting from either confinement (*46*) or friction (*52*).

This theory predicts a phase diagram composed of three distinct regimes (**Fig. 2f**):

i) the active sheet buckles out-of-plane because only out-of-plane modes are unstable: $0 < \zeta/\mu < 1$;

ii) the active sheet buckles out-of-plane because the most unstable out-of-plane mode grows faster than the most-unstable in-plane mode: $1 < \zeta/\mu < \frac{1}{1-(1+\frac{\kappa}{K})^{-0.5}}$ where $\kappa$ is the bending modulus and K is the nematic elasticity;

iii) the active sheet bends and stretches in-plane because the most unstable in-plane mode grows faster than the most-unstable out-of-plane mode: $\zeta/\mu > \frac{1}{1-(1+\frac{\kappa}{K})^{-0.5}}$

This minimal model demonstrates that the direction of the instability is set by the balance between active stresses and passive elastic stresses. This continuum description is consistent with our experimental observations and demonstrates that a thin active crosslinked network can either exhibit a fluid-like bend instability or a solid-like out-of-plane buckling instability.

**A reaction kinetics model connects molecular composition to the activity and mechanical properties of the network.**

We developed a model based on Michaelis-Menten enzyme kinetics to connect the concentration of cytoskeletal proteins to the macroscopic material parameters that determine the dynamics of the active network. Motor proteins can either generate forces or act as a crosslinker depending on ATP concentration. Here, this simple model estimates the number of active motors (ATP-bound motors stepping on microtubules) and the number of passive motors (non-ATP-bound motors that passively crosslink the network, see SI for details). Extensile stresses are generated by the relative sliding of microtubules past one another (*2*). Therefore, we assumed that the activity $\zeta$ is proportional to the elongation rate of a microtubule bundle, which itself depends linearly on the number of ATP-bound motors clusters sliding the microtubules by the Michaelis-Menten relation (See SI, section 2.1):

$$\zeta = \zeta_0 . \frac{[ATP].[Motor\ clusters]}{\frac{k_h + k_-}{k_b} + [ATP]}$$

where $\zeta_0$ is an activity constant that depends on the efficiency with which ATP hydrolysis translate into mechanical work, and $k_h$ is the ATP hydrolysis rate, $k_-$ and $k_b$ are respectively the unbinding and binding rates of ATP to kinesin motors (*53*). Next, bulk rheology experiments on passive networks of crosslinked biopolymers showed that the shear modulus μ and the bending modulus $\kappa$ vary as $\mu = \mu_0[crosslinkers]^2$ and $\kappa = \kappa_0[crosslinkers]^2$ (see SI section 2) (*49, 54*). Finally, we estimated the total number of crosslinkers as follow:

$$[crosslinkers] = p_0[PRC1] + [motor\ clusters]_{total} - [motor\ clusters]_{stepping}$$

PRC1 being a sensitive protein to purify, we considered that only a fraction $p_0$ of the proteins is active. The last two terms on the right-hand side correspond to the fraction of molecular motors that crosslink microtubules instead of sliding them apart.

Combining the hydrodynamic model with the enzyme kinetics, we derived a theory with four unknown parameters: i) the ratio of the activity $\zeta_0$ and $\mu_0$, ii) the ratio of the nematic elasticity $K_0$ and $\kappa_0$, iii) the ratio of $K_0$ and $\zeta_0$, and iv) $p_0$ (see SI, section 2.4 for the equation of the phase boundary). We performed over N=1000 experiments to build four experimental phase diagrams showing how the transition from an active fluid to an active solid depends on ATP, motor clusters, and crosslinker concentrations, and the average length of the microtubules (**Fig. 3**).

**Fig. 3: Biochemical processes control the transition from an active fluid to an active solid.**

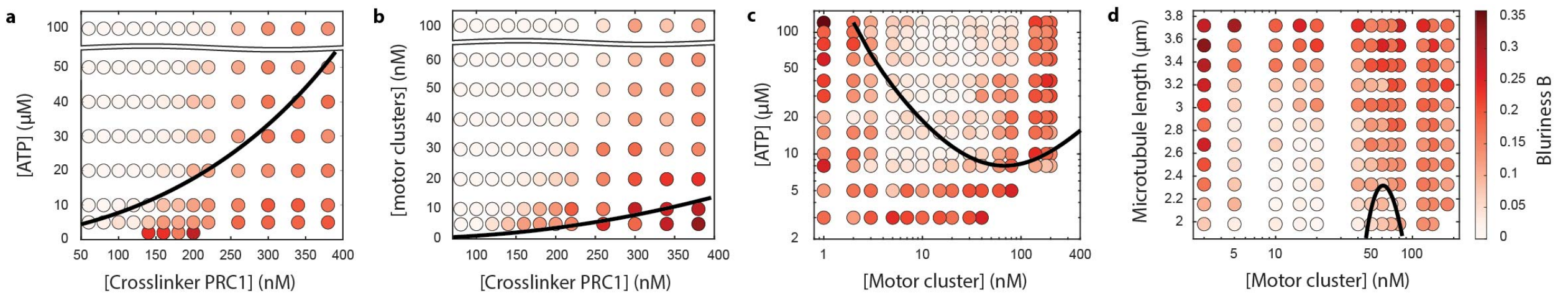

Experimental phase diagrams and theoretical phase boundaries (continuous black lines) describing the transition from an in-plane fluid-like bend instability to an the out-of-plane solid-like buckling for various a) ATP and PRC1 crosslinker concentrations ([motor clusters]=60nM), b) motor clusters and PRC1 concentrations ([ATP]=1mM), c) ATP and motor clusters concentrations ([PRC1]= 100nM), and d) microtubules' length and motor cluster concentration ([ATP]=8µM, [PRC1]=100nM). The color map indicates the blurriness B and is the same for the four phase diagrams. Each colored circle corresponds to a distinct experimental realization.

The theoretical model recovers that the more crosslinked the network is, the more activity is required to fluidize the network (**Figs. 3a-b**). The model also captures the re-entrant transition controlled by motor clusters concentration: increasing the number of motors first fluidizes but then induces a rigidification of the network as adding more motors in the ATP limiting case is equivalent to increasing the number of crosslinkers (**Fig. 3c**). Finally, the model includes the influence of the length of the microtubules through its impact on the nematic elastic cost, which qualitatively describes the transition from an active fluid for short microtubules to an active solid for longer microtubules (**Fig. 3d**). The out-of-plane wavelengths are also well captured by the theory (**Fig. S5**).

While this model describes well the re-entrant phases, the range of motor cluster concentrations where the fluid phase is observed is slightly different in the experiments and in the theory ([motor cluster]$_{exp}$=20nM while [motor cluster]$_{th}$=70nM for the extrema of the fluid-solid phase boundary in **Fig. 3c-d**). This discrepancy might result from an ambiguity of the proper scaling for the elasticity of a nematic elastomer composed of microtubules (see SI section 2.2). Of note, the ratio of motors to crosslinkers required to fluidize the network is around 10%. In the regime explored here, each microtubule is decorated on average by 0.2-25 motor clusters and 10-80 crosslinkers (*55*). The quantitative agreement between theory and experimental data provides a multiscale map between the activity and shear modulus of the network and its molecular composition (**Fig. 4,** see SI section 3 for fitting procedure). Interestingly, increasing ATP induces a softening of the network. A similar effect has been observed for the nematic elasticity of a microtubule-based 2D active nematic liquid crystal (*23, 56*).

**Fig. 4: Molecular composition of the network sets its activity and elasticity.**

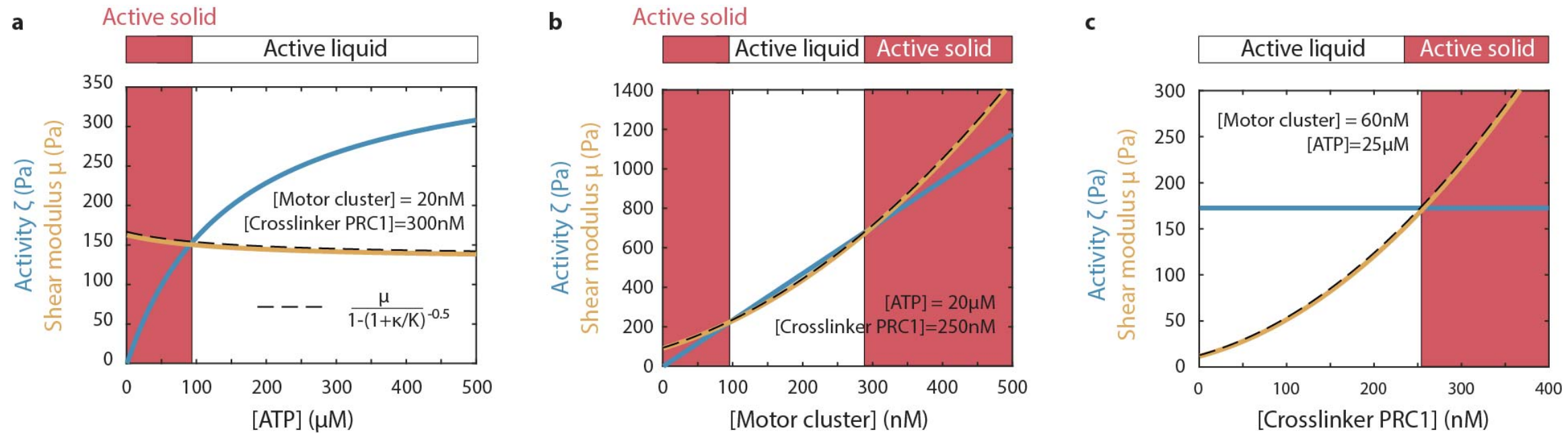

Activity (blue curves) and shear modulus (orange curves) when varying a) ATP concentration, b) motor clusters concentrations, and c) PRC1 crosslinkers concentrations. Active and elastic stresses are inferred from fitting the theoretical model to the experimental phase diagrams (as described in the SI, sections 2.4 and 3). When the activity is above (resp. below) the dashed line, the network deforms like an active fluid (resp. active solid). $\mu_0$ is estimated from bulk rheology experiments by Gagnon et al (*35*) (see SI section 3.3), while the ratios $\zeta_0/\mu_0$, $K_0/\kappa_0$ and, $\kappa_0/\zeta_0$ are fitted parameters.

**Fig. 5: Optogenetic control of the direction of the instability.**

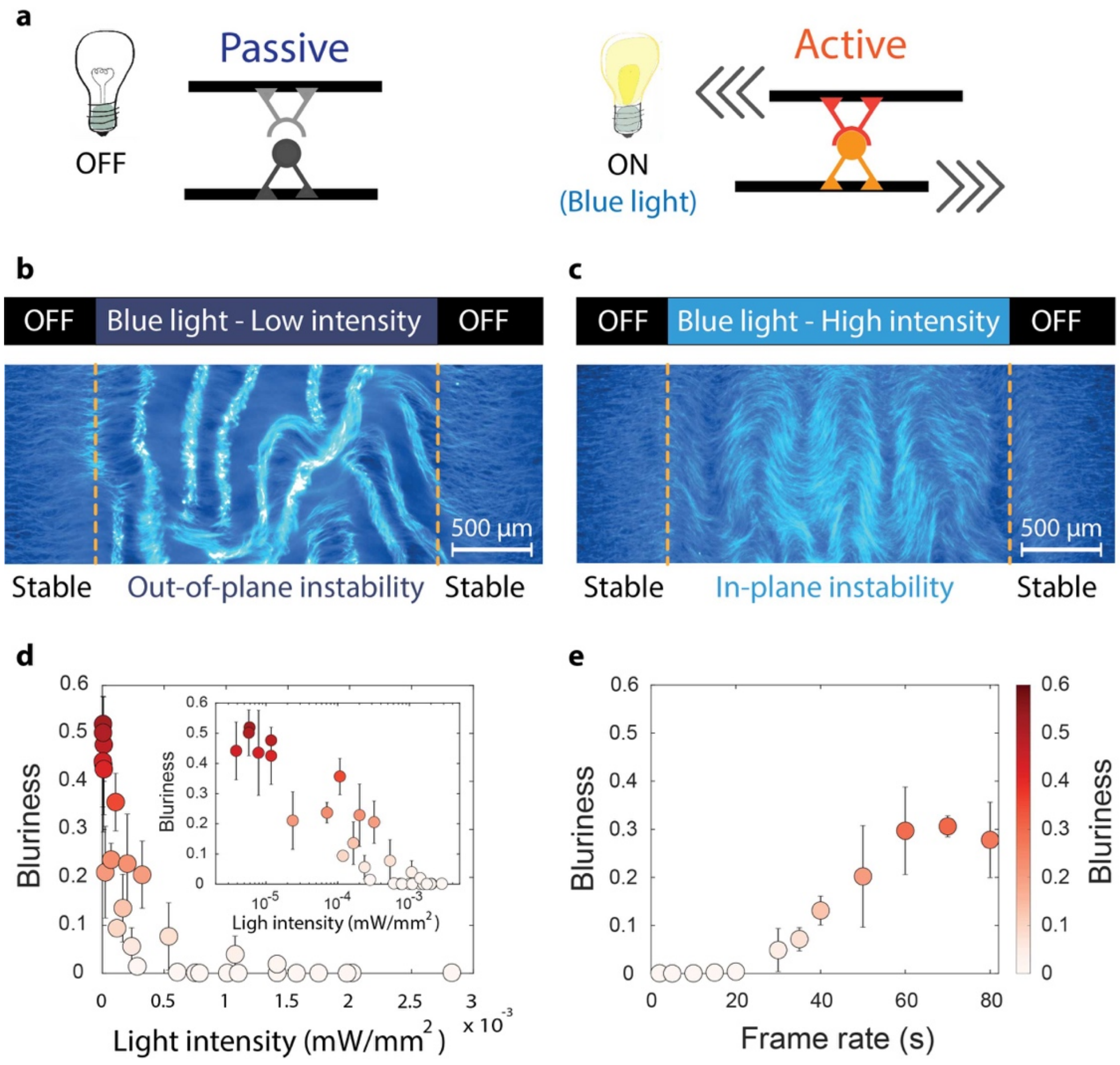

a) Schematics for the light-dimerizable kinesin motors. In the absence of light, the motors hydrolyze ATP, step on individual microtubules, but do not dimerize, which does not induce any relative sliding of the microtubules. b) When blue light is turned on (488nm wavelength), the motors dimerize. Their stepping produces a net extension of the microtubule bundle. We shined pulses patterns of blue light on a flow-aligned network; b) for low blue light intensity, the network buckles out-of-plane like an active solid; c) for high blue light intensity, the network bends in-plane like an active fluid. In the absence of light, the network is stable and stays flow aligned; d) Blurriness B of the unstable active network for varying light intensity (main: lin-lin plot, insert: lin-log plot showing the sharp transition around 0.1 $\mu W/mm^3$). e) Increasing the blue light exposure rate induces a transition from in-plane bending to out-of-plane buckling for a critical exposure rate $\tau \sim 30\ sec$. The color map is the same for d) and e) and indicates the blurriness B.

**Optogenetic control of the molecular motors sets the direction of the most unstable mode.**

We further demonstrate how to leverage light-activable molecular motors to optogenetically control the activity and elasticity of the network *in situ*. We replaced the conventional motor clusters with light-dimerizable kinesin motors (*27, 57*). In the absence of light, the motors do not dimerize: they hydrolyze ATP, step onto microtubules, but do not induce any relative sliding (**Fig.**

**5a**). When exposed to blue light, the motors form a cluster that can slide apart adjacent microtubules (**Fig. 5b**), much like the extension produced by the conventional kinesin clusters. These light-dimerizable motors enable spatiotemporal control of motor activity. In particular, we show that tuning the intensity and the frequency of activity pulses modulate the direction of the instability. First, the network was stable in the absence of light. Then, at low intensity, the network buckled out-of-plane (**Fig. 5c**). Increasing the light intensity induced a transition to an in-plane instability (**Fig. 5d**), which is consistent with the number of dimerized motors increasing and then saturating with the light intensity. The wavelengths of the instability were also consistent with an increase in activity (**Fig. S6**). The direction of the instability can also be controlled by changing the frequency of the light pulses (**Fig. 5e**). Shining high-intensity light at a fast frame rate triggered an in-plane instability while shining light at a slow frame rate triggered an out-of-plane buckling. The critical pulse rate required to fluidize the network is around 30 sec. While this transition cannot be captured by the quasistatic approximation of our model, it is consistent with the off-rate of the light-induced dimer $\tau \sim 30\ sec$ (*58*): if the framerate is larger than 30 sec, then only a fraction of the stepping motors is dimerized, which result in a lower activity. Finally, we note that deformations only partially relax once the activity is turned off, probably due to the presence of crosslinkers (**Fig. S7, Video S8**). In the future, other control strategies, maybe with light-controllable crosslinkers, could be implemented to reversibly deform microtubule-based active materials in 3D.

**Discussion**

To fully harness the engineering potential of active matter and trigger self-organization on demand, one needs to be able to both a) design materials with targeted activity and mechanical properties, and b) control these properties in space and time. Here we have achieved both. First, we predictively designed coarse-grained active and elastic stresses from microscopic building blocks by combined quantitative measurements with theory at two levels – hydrodynamics and enzyme kinematics. While simplifications in the Michaelis-Menten kinetics neglect some of the molecular complexity of the cytoskeletal proteins, like the load dependance of kinesin stepping (*53, 59*), this work leads the way to rationally design active materials with targeted mechanical properties.

Second, we identified and implemented spatiotemporal design principles to drive targeted deformations in 3D, controlling the active and elastic stresses *in situ.* Previously, control of active matter focused mostly on 2D dynamics and relied on tuning the effective motility of topological defects in 2D active nematics (*25*) and in 2D polar active matter (*27*). From a material science perspective, our work yields insights into developing reconfigurable active materials where transitions between non-equilibrium states can be triggered spontaneously or externally. From a life science perspective, this framework illustrates how cytoskeletal mechanics is controlled via spatiotemporal activity patterns. More generally, this work might shed light on how the spatiotemporal regulation of a finite pool of proteins can fluidize or rigidify the cell cytoplasm to drive collective flows while transmitting forces within a cohesive tissue *in vivo*. For example, during gastrulation, cells in the ventral tissue coordinate pulses of actomyosin contractility to drive a solid-to-fluid transition, triggering the out-of-plane deformation of a 2D tissue (*60-62*). Contrarily, during vertebrate body axis elongation (*63*), tissues undergo a timed fluid-to-solid jamming transition that promotes tail bud formation. Our work suggests that the spatiotemporal control of the activity and the mechanical properties of the cytoskeleton is a generic physical mechanism that governs how living and active matter change shape in 3D.

**Acknowledgements**
We thank the director of the Brandeis Biomaterial Facility Shibani Dalal, for help with protein purification. We thank Tyler Ross for help with the opto-K365 constructs. We thank Guillaume Sarfati, Jean-Christophe Galas, Andre Estevez-Torres, Hannah Yevick, Peter Foster, Zvonimir Dogic, Michael Hagan, and John Edison for discussion.
*Note:* In the concluding stages of our work, we became aware of a complementary, independent effort by the groups of Andre Estevez-Torres and Jean-Christophe Galas who studied instabilities in active fluids composed of microtubule bundled by a depletant (*64*).

**Funding**
B.N. and G.D. acknowledge support from a NSF CAREER award DMR-2047119. We also acknowledge the use of the optical, microfluidics, and biomaterial facilities supported by NSF MRSEC DMR-2011846. M.V., L.S., L.L., A.B. and G.D. acknowledge support of NSF MRSEC DMR-2011846. Computational resources were provided by the Brandeis HPCC which is partially supported by the NSF through DMR-MRSEC 2011846 and OAC-1920147.

**Author contributions**
G.D. designed the research; B.N. performed the experiments; B.N. and G.D. analyzed the data; M.V. and A.B. developed the theoretical model; M.V., B.N., and L.T. fitted the data; L.L. contributed to the experiments with optogenetic molecular motors; G.D. and A.B. supervised the research; B.N., M.V., A.B, and G.D. wrote the manuscript.

**References**

1. M. C. Marchetti, J.-F. Joanny, S. Ramaswamy, T. B. Liverpool, J. Prost, M. Rao, R. A. Simha, Hydrodynamics of soft active matter. *Reviews of Modern Physics* **85**, 1143 (2013).
2. T. Sanchez, D. T. Chen, S. J. DeCamp, M. Heymann, Z. Dogic, Spontaneous motion in hierarchically assembled active matter. *Nature* **491**, 431-434 (2012).
3. J. Dunkel, S. Heidenreich, K. Drescher, H. H. Wensink, M. Bar, R. E. Goldstein, Fluid dynamics of bacterial turbulence. *Phys. Rev. Lett.* **110**, 228102 (2013).
4. S. Zhou, A. Sokolov, O. D. Lavrentovich, I. S. Aranson, Living liquid crystals. *Proc. Natl. Acad. Sci. U. S. A.* **111**, 1265-1270 (2014).
5. A. Bricard, J.-B. Caussin, N. Desreumaux, O. Dauchot, D. Bartolo, Emergence of macroscopic directed motion in populations of motile colloids. *Nature* **503**, 95-98 (2013).
6. M. Driscoll, B. Delmotte, M. Youssef, S. Sacanna, A. Donev, P. Chaikin, Unstable fronts and motile structures formed by microrollers. *Nature Physics* **13**, 375-379 (2016).
7. V. Soni, E. S. Bililign, S. Magkiriadou, S. Sacanna, D. Bartolo, M. J. Shelley, W. T. M. Irvine, The odd free surface flows of a colloidal chiral fluid. *Nature Physics* **15**, 1188-1194 (2019).

8. S. Banerjee, M. L. Gardel, U. S. Schwarz, The Actin Cytoskeleton as an Active Adaptive Material. *Annu Rev Condens Matter Phys* **11**, 421-439 (2020).
9. L. R. Serbus, B. J. Cha, W. E. Theurkauf, W. M. Saxton, Dynein and the actin cytoskeleton control kinesin-driven cytoplasmic streaming in Drosophila oocytes. *Development* **132**, 3743-3752 (2005).
10. C. L. Rieder, A. Khodjakov, Mitosis Through the Microscope: Advances in Seeing Inside Live Dividing Cells. *Science* **300**, (2003).
11. H. Wioland, F. G. Woodhouse, J. Dunkel, J. O. Kessler, R. E. Goldstein, Confinement stabilizes a bacterial suspension into a spiral vortex. *Phys. Rev. Lett.* **110**, 268102 (2013).
12. K.-T. Wu, J. B. Hishamunda, D. T. N. Chen, S. J. DeCamp, Y.-W. Chang, A. Fernández-Nieves, S. Fraden, Z. Dogic, Transition from turbulent to coherent flows in confined three-dimensional active fluids. *Science* **355**, eaal1979 (2017).
13. F. J. Nedelec, T. Surrey, A. C. Maggs, S. Leibler, Self-organization of microtubules and motors. *Nature* **389**, 305–308 (1997).
14. K. Kruse, J. F. Joanny, F. Julicher, J. Prost, K. Sekimoto, Asters, vortices, and rotating spirals in active gels of polar filaments. *Phys Rev Lett* **92**, 078101 (2004).
15. F. Julicher, S. W. Grill, G. Salbreux, Hydrodynamic theory of active matter. *Rep Prog Phys* **81**, 076601 (2018).
16. S. Furthauer, B. Lemma, P. J. Foster, S. C. Ems-McClung, C. H. Yu, C. E. Walczak, Z. Dogic, D. J. Needleman, M. J. Shelley, Self-straining of actively crosslinked microtubule networks. *Nat Phys* **15**, 1295-1300 (2019).
17. T. B. Liverpool, M. C. Marchetti, Instabilities of isotropic solutions of active polar filaments. *Phys. Rev. Lett.* **90**, 138102 (2003).
18. T. B. Liverpool, M. C. Marchetti, Rheology of active filament solutions. *Phys Rev Lett* **97**, 268101 (2006).
19. T. Gao, R. Blackwell, M. A. Glaser, M. D. Betterton, M. J. Shelley, Multiscale polar theory of microtubule and motor-protein assemblies. *Phys. Rev. Lett.* **114**, 048101 (2015).
20. M. S. D. Wykes, J. Palacci, T. Adachi, L. Ristroph, X. Zhong, M. D. Ward, J. Zhang, M. J. Shelley, Dynamic self-assembly of microscale rotors and swimmers. *Soft matter* **12**, 4584-4589 (2016).
21. J. M. Moore, T. N. Thompson, M. A. Glaser, M. D. Betterton, Collective motion of driven semiflexible filaments tuned by soft repulsion and stiffness. *Soft Matter* **16**, 9436-9442 (2020).
22. J. Alvarado, M. Sheinman, A. Sharma, F. C. MacKintosh, G. H. Koenderink, Molecular motors robustly drive active gels to a critically connected state. *Nature Physics* **9**, 591-597 (2013).
23. J. Colen, M. Han, R. Zhang, S. A. Redford, L. M. Lemma, L. Morgan, P. V. Ruijgrok, R. Adkins, Z. Bryant, Z. Dogic, M. L. Gardel, J. J. de Pablo, V. Vitelli, Machine learning active-nematic hydrodynamics. *Proc Natl Acad Sci U S A* **118**, (2021).
24. Z. Zhou, C. Joshi, R. Liu, M. M. Norton, L. Lemma, Z. Dogic, M. F. Hagan, S. Fraden, P. Hong, Machine learning forecasting of active nematics. *Soft Matter* **17**, 738-747 (2021).
25. R. Zhang, S. A. Redford, P. V. Ruijgrok, N. Kumar, A. Mozaffari, S. Zemsky, A. R. Dinner, V. Vitelli, Z. Bryant, M. L. Gardel, J. J. de Pablo, Spatiotemporal control of liquid crystal structure and dynamics through activity patterning. *Nat Mater* **20**, 875-882 (2021).

26. T. Nitta, Y. Wang, Z. Du, K. Morishima, Y. Hiratsuka, A printable active network actuator built from an engineered biomolecular motor. *Nat Mater* **20**, 1149-1155 (2021).
27. T. D. Ross, H. J. Lee, Z. Qu, R. A. Banks, R. Phillips, M. Thomson, Controlling organization and forces in active matter through optically defined boundaries. *Nature* **572**, 224-229 (2019).
28. M. M. Norton, P. Grover, M. F. Hagan, S. Fraden, Optimal Control of Active Nematics. *Phys Rev Lett* **125**, 178005 (2020).
29. C. G. Wagner, M. M. Norton, J. S. Park, P. Grover, Exact coherent structures and phase space geometry of pre-turbulent 2D active nematic channel. *arXiv:2109.06455v1*, (2021).
30. S. Liu, S. Shankar, M. C. Marchetti, Y. Wu, Viscoelastic control of spatiotemporal order in bacterial active matter. *Nature* **590**, 80-84 (2021).
31. R. A. Simha, S. Ramaswamy, Hydrodynamic fluctuations and instabilities in ordered suspensions of self-propelled particles. *Phys. Rev. Lett.* **89**, 058101 (2002).
32. A. Senoussi, S. Kashida, R. Voituriez, J. C. Galas, A. Maitra, A. Estevez-Torres, Tunable corrugated patterns in an active nematic sheet. *Proc. Natl. Acad. Sci. U. S. A.* **116**, 22464-22470 (2019).
33. T. Strubing, A. Khosravanizadeh, A. Vilfan, E. Bodenschatz, R. Golestanian, I. Guido, Wrinkling Instability in 3D Active Nematics. *Nano Lett* **20**, 6281-6288 (2020).
34. D. Humphrey, C. Duggan, D. Saha, D. Smith, J. Kas, Active fluidization of polymer networks through molecular motors. *Nature* **416**, (2002).
35. D. A. Gagnon, C. Dessi, J. P. Berezney, R. Boros, D. T. N. Chen, Z. Dogic, D. L. Blair, Shear-Induced Gelation of Self-Yielding Active Networks. *Phys. Rev. Lett.* **125**, (2020).
36. S. Ramaswamy, Active fluids. *Nature Reviews Physics* **1**, 640-642 (2019).
37. S. J. DeCamp, G. S. Redner, A. Baskaran, M. F. Hagan, Z. Dogic, Orientational order of motile defects in active nematics. *Nature Materials* **14**, 1110 (2015).
38. G. Duclos, R. Adkins, D. Banerjee, M. S. E. Peterson, M. Varghese, I. Kolvin, A. Baskaran, R. A. Pelcovits, T. R. Powers, A. Baskaran, F. Toschi, M. F. Hagan, S. J. Streichan, V. Vitelli, D. A. Beller, Z. Dogic, Topological structure and dynamics of three-dimensional active nematics. *Science* **367**, 1120-1124 (2020).
39. R. Subramanian, E. M. Wilson-Kubalek, C. P. Arthur, M. J. Bick, E. A. Campbell, S. A. Darst, R. A. Milligan, T. M. Kapoor, Insights into antiparallel microtubule crosslinking by PRC1, a conserved nonmotor microtubule binding protein. *Cell* **142**, 433-443 (2010).
40. P. Chandrakar, J. Berezney, B. Lemma, B. Hishamunda, A. Berry, K.-T. Wu, R. Subramanian, J. Chung, D. Needleman, J. Gelles, Microtubule-based active fluids with improved lifetime, temporal stability and miscibility with passive soft materials. *arXiv preprint arXiv:1811.05026*, (2018).
41. A. J. Tan, E. Roberts, S. A. Smith, U. A. Olvera, J. Arteaga, S. Fortini, K. A. Mitchell, L. S. Hirst, Topological chaos in active nematics. *Nature Physics* **15**, 1033-1039 (2019).
42. B. Martínez-Prat, R. Alert, F. Meng, J. Ignés-Mullol, J.-F. Joanny, J. Casademunt, R. Golestanian, F. Sagués, Scaling Regimes of Active Turbulence with External Dissipation. *Physical Review X* **11**, (2021).
43. R. Alert, J.-F. Joanny, J. Casademunt, Universal scaling of active nematic turbulence. *Nature Physics*, (2020).

44. C. Lang, J. Kohlbrecher, L. Porcar, A. Radulescu, K. Sellinghoff, J. K. G. Dhont, M. P. Lettinga, Microstructural Understanding of the Length-and Stiffness-Dependent Shear Thinning in Semidilute Colloidal Rods. *Macromolecules*, (2019).
45. T. A. J. Lenstra, Z. Dogic, J. K. G. Dhont, Shear-induced displacement of isotropic-nematic spinodals. *The Journal of Chemical Physics* **114**, 10151-10162 (2001).
46. P. Chandrakar, M. Varghese, S. A. Aghvami, A. Baskaran, Z. Dogic, G. Duclos, Confinement Controls the Bend Instability of Three-Dimensional Active Liquid Crystals. *Phys Rev Lett* **125**, 257801 (2020).
47. W. R. Schief, R. H. Clark, A. H. Crevenna, J. Howard, Inhibition of kinesin motility by ADP and phosphate supports a hand-over-hand mechanism. *Proc Natl Acad Sci U S A* **101**, 1183-1188 (2004).
48. L. Pocivavsek, R. Dellsy, A. Kern, S. Johnson, B. Lin, K. Y. Lee, E. Cerda, Stress and fold localization in thin elastic membranes. *Science* **320**, 912-916 (2008).
49. J. H. Shin, M. L. Gardel, L. Mahadevan, P. Matsudaira, D. A. Weitz, Relating microstructure to rheology of a bundled and cross-linked F-actin network in vitro. *Proc Natl Acad Sci U S A* **101**, 9636-9641 (2004).
50. J. Prost, F. Jülicher, J. F. Joanny, Active gel physics. *Nature Physics* **11**, 111-117 (2015).
51. J. Rønning, M. C. Marchetti, M. J. Bowick, L. Angheluta, Flow around topological defects in active nematic films. *ArXiv preprint arXiv:2111.08537v1*, (2021).
52. B. Martínez-Prat, J. Ignés-Mullol, J. Casademunt, F. Sagués, Selection mechanism at the onset of active turbulence. *Nature Physics* **15**, 362-366 (2019).
53. K. Visscher, M. J. Schnitzer, S. M. Block, Single kinesin molecules studied with a molecular force clamp. *Nature* **400**, (1999).
54. M. L. Gardel, J. H. Shin, F. C. MacKintosh, L. Mahadevan, P. Matsudaira, D. A. Weitz, Elastic Behavior of Cross-Linked and Bundled Actin Networks. *Science* **304**, (2004).
55. S. F□urthauer, M. J. Shelley, How crosslink numbers shape the large-scale physics of cytoskeletal materials. *arXiv:2106.13273v1*, (2021).
56. L. M. L. S. J. D. Z. Y. L. G. Z. Dogic, Statistical Properties of Autonomous Flows in 2D Active Nematics. *Soft matter* **15**, 3264-3272 (2019).
57. L. M. Lemma, T. D. Ross, Z. Dogic, in preparation.
58. G. Guntas, R. A. Hallett, S. P. Zimmerman, T. Williams, H. Yumerefendi, J. E. Bear, B. Kuhlman, Engineering an improved light-induced dimer (iLID) for controlling the localization and activity of signaling proteins. *Proc Natl Acad Sci U S A* **112**, 112-117 (2015).
59. H. Khataee, J. Howard, Force Generated by Two Kinesin Motors Depends on the Load Direction and Intermolecular Coupling. *Phys Rev Lett* **122**, 188101 (2019).
60. A. C. Martin, M. Kaschube, E. F. Wieschaus, Pulsed contractions of an actin-myosin network drive apical constriction. *Nature* **457**, 495-499 (2009).
61. J. Solon, A. Kaya-Copur, J. Colombelli, D. Brunner, Pulsed forces timed by a ratchet-like mechanism drive directed tissue movement during dorsal closure. *Cell* **137**, 1331-1342 (2009).
62. L. Atia, D. Bi, Y. Sharma, J. A. Mitchel, B. Gweon, S. Koehler, S. J. DeCamp, B. Lan, J. H. Kim, R. Hirsch, A. F. Pegoraro, K. H. Lee, J. R. Starr, D. A. Weitz, A. C. Martin, J. A. Park, J. P. Butler, J. J. Fredberg, Geometric constraints during epithelial jamming. *Nat Phys* **14**, 613-620 (2018).

63. A. Mongera, P. Rowghanian, H. J. Gustafson, E. Shelton, D. A. Kealhofer, E. K. Carn, F. Serwane, A. A. Lucio, J. Giammona, O. Campas, A fluid-to-solid jamming transition underlies vertebrate body axis elongation. *Nature* **561**, 401-405 (2018).
64. G. Sarfati, A. Maitra, R. Voituriez, J.-C. Galas, A. Estevez-Torres, Crosslinking and depletion determine spatial instabilities in cytoskeletal active matter. *arXiv preprint arXiv:2112.11361v1* (2021).

Supplementary Information

# Dual antagonistic role of motor proteins in fluidizing active networks

Bibi Najma[1,*], Minu Varghese[1,2,*], Lev Tsidilkovski[1], Linnea Lemma[1,3,4], Aparna Baskaran[1], and Guillaume Duclos[1]

[1]Department of Physics, Brandeis University, Waltham, MA 02453
[2]Department of Physics, University of Michigan, Ann Arbor, MI 48109
[3]Department of Physics, University of California at Santa Barbara, Santa Barbara, CA 93106
[4]Current address: Department of Chemical and Biological Engineering, Princeton University, Princeton, NJ 08544
[*]These authors contributed equally

## Contents

# 1 Hydrodynamic Theory

Consider a collection of rod-like units (microtubule filaments in the experiments) in the xy plane, all of which are initially aligned along the x-axis. Due to excluded volume interactions, there is a Frank free energy associated with distortions about this nematically ordered state, given by [1]

$$\mathscr{F}_{\text{nematic}} = \frac{1}{2}\int dx\ dy\ K\ |\vec{\nabla}\hat{n}|^2 \tag{1}$$

where $K$ is the Frank elastic constant and $\hat{n}$ is the nematic director which represents the local orientation of the rod-like units, with $\hat{n}(t=0) = \hat{n}_0 = \hat{x}$. Suppose that the rod-like units are cross-linked while in this configuration (by PRC1 or KSA in the experiments). The cross-linking turns the collection of rods into an elastic sheet whose undeformed state corresponds to the initial configuration of the system. Let the initial configuration of the sheet (i.e., the position of material points on the sheet) be $\vec{R}_0(x,y) = x\hat{x} + y\hat{y}$, and its current configuration be $\vec{R}(x,y) = \vec{R}_0(x,y) + \vec{u}(x)$, where $\vec{u} = u_x\hat{x} + u_y\hat{y} + h\hat{z}$ is assumed to be small. We have made the simplifying assumption of $\vec{u}$ being purely a function of $x$ based on experimental observations. The elastic cost of the deformation,

$$\mathscr{F}_{\text{elastic}} = \frac{1}{2}\int dxdy\ \ \left[\nu(\partial_x u_x)^2 + \mu(\partial_x u_y)^2 + \kappa(\partial_x^2 h)^2\right] \tag{2}$$

where $\mu$ is the shear modulus, $\nu$ is a modified bulk modulus, and $\kappa$ is a modified bending modulus (modified from their isotropic values due to the presence of nematic order).

The nematic director in the deformed state is $n^\alpha = \frac{\partial R^\alpha}{\partial R_0^\beta} n_0^\beta$, or $\hat{n} = \hat{n}_0 + \hat{n}_0.\vec{\nabla}\vec{u} = (1+\partial_x u_x)\ \hat{x} + \partial_x u_y\ \hat{y} + \partial_x h\ \hat{z}$. Thus, the total free energy associated with a material deformation is

$$\mathscr{F}[u_x, u_y, h] = \mathscr{F}_{\text{nematic}} + \mathscr{F}_{\text{elastic}} \tag{3}$$

$$= \frac{1}{2}\int dxdy\ \ \left[\nu(\partial_x u_x)^2 + \mu(\partial_x u_y)^2 + \kappa(\partial_x^2 h)^2 + K[(\partial_x^2 u_y)^2 + (\partial_x^2 h)^2]\right] \tag{4}$$

Now, suppose that the material has extensile activity (resulting from ATP driven kinesin motors sliding microtubule bundles in the experiments). This results in an active force [2],

$$\vec{f} = -\zeta\vec{\nabla}\cdot(\vec{n}\vec{n}) = -\zeta(\partial_x^2 u_y\hat{y} + \partial_x^2 h\hat{z}) \tag{5}$$

where $\zeta > 0$ is proportional to the rate at which the microtubules slide past each other. Assuming a viscous drag between the membrane and the (quasistatic) ambient medium, the rate of energy dissipation due to friction is

$$\mathscr{R} = \frac{\gamma}{2}\int dxdy\ (\partial_t\vec{u})^2 \tag{6}$$

Balancing the conservative, active, and frictional forces,

$$\frac{\delta \mathscr{F}}{\delta u_i} + \frac{\partial \mathscr{R}}{\partial \dot{u}_i} + f_i = 0 \tag{7}$$

i.e.,

$$\partial_t u_x = \frac{\nu}{\gamma}\partial_x^2 u_x, \quad \partial_t u_y = \frac{1}{\gamma}\left[(\mu-\zeta)\partial_x^2 - K\partial_x^4\right]u_y, \quad \partial_t h = -\frac{1}{\gamma}\left[(K+\kappa)\partial_x^4 + \zeta\partial_x^2\right]h \tag{8}$$

In Fourier space,

$$\partial_t \tilde{u}_x = -\frac{\nu}{\gamma}q_x^2 \tilde{u}_x, \quad \partial_t \tilde{u}_y = \frac{1}{\gamma}\left[(\zeta-\mu)q_x^2 - Kq_x^4\right]\tilde{u}_y, \quad \partial_t \tilde{h} = \frac{1}{\gamma}\left[\zeta q_x^2 - (K+\kappa)q_x^4\right]\tilde{h} \tag{9}$$

**In-plane instability** The growth rate of an in-plane instability with wavenumber $q_x$ is $\frac{1}{\gamma}\left[(\zeta-\mu)q_x^2 - Kq_x^4\right]$. Note that this expression is negative for all $q_x$ (i.e., there is no in-plane instability) if $\zeta < \mu$, while it is positive for $q_x < \sqrt{\frac{\zeta-\mu}{K}}$ when $\zeta > \mu$. The fastest growing wave mode corresponds to $q_x = \sqrt{\frac{\zeta-\mu}{2K}}$, so that the wavelength observed in experiments should be $2\pi\sqrt{\frac{2K}{\zeta-\mu}}$. When the in plane instability exists, the growth rate of the fastest growing mode is $\frac{(\zeta-\mu)^2}{4\gamma K}$

**Out-of-plane instability** The growth rate of an out-of-plane instability of wavenumber $q_x$ is $-\frac{1}{\gamma}\left[(K+\kappa)q_x^4 - \zeta q_x^2\right]$. This expression is always positive for $q_x < \sqrt{\frac{\zeta}{K+\kappa}}$. The fastest growing wavemode corresponds to $q_x = \sqrt{\frac{\zeta}{2(K+\kappa)}}$, so the experimentally observed out-of-plane wavelength should be $2\pi\sqrt{\frac{2(K+\kappa)}{\zeta}}$. The growth rate of the fastest growing mode is $\frac{\zeta^2}{4\gamma(K+\kappa)}$

**Transition from out of plane to in plane instability** At small activities ($\zeta < \mu$), the instability is purely out of plane. For $\zeta > \mu$, in-plane instability has a growth rate of $\frac{(\zeta-\mu)^2}{4\gamma K}$, while out of plane instability has a growth rate of $\frac{\zeta^2}{4\gamma(K+\kappa)}$. Thus, the transition from out of plane to in plane instability happens when $\left(1-\frac{\mu}{\zeta}\right)^2(1+\frac{\kappa}{K}) - 1 = 0$, or $\frac{\zeta}{\mu} = \frac{1}{1-\frac{1}{\sqrt{1+\frac{\kappa}{K}}}}$

# 2 Mapping between experimental and theoretical parameters

In this section, we estimate the phenomenological constants that appear in the hydrodynamic theory based on simple phenomenological models.

## 2.1 Activity

The chemical reaction that generates extensile activity corresponds to hydrolysis of an ATP molecule that is bound to a kinesin motor cluster (called KSA in what follows)[3]

$$\text{ATP+KSA} \underset{k_-}{\overset{k_b}{\rightleftharpoons}} \text{ATP.KSA} \xrightarrow{k_h} \text{ADP.KSA + P} \tag{10}$$

where $k_b = 2\mu M^{-1}s^{-1}$ [4, 5], $k_- = 200s^{-1}$ and $k_h = 100s^{-1}$[4, 5, 6, 7]. We have assumed there is an excess of microtubules/tubulin compared to KSA( [tubulin]=13.3uM, [KSA]$\leq$ 400nM), so that almost all kinesin motors are attached to microtubules. In what follows, we will also assume that there is an excess of ATP compared to KSA (1$\leq$[ATP]$\leq$ 100$\mu$M), so that almost all the ATP in the system exist in their free state (rather than bound to KSA). Both these assumptions hold very well for the experimental system, since it is prepared with micro molar concentrations of tubulin and ATP, compared to nano molar concentration of KSA. There is an additional step in the ATP cycle, where ADP is converted to ATP by the action of an enzyme. However, this reaction is so fast that there is very little ADP in the system at any time. The rate of change in concentration of the ATP. KSA complex is given by

$$\frac{d}{dt}[\text{ATP.KSA}] = k_b[\text{ATP}][\text{KSA}]_0 - k_h[\text{ATP.KSA}] - k_-[\text{ATP.KSA}] \tag{11}$$

where $[\text{KSA}]_0$ represents the concentration of KSA that is not attached to ATP, and is given by $[\text{KSA}]_0$=[KSA]−[ATP.KSA] where [KSA] is the amount of KSA used to prepare the solution, and we have made the approximation $[\text{ATP}]_0 \approx$[ATP]. At steady state,

$$k_b[\text{ATP}][\text{KSA}]_0 = (k_h + k_-)[\text{ATP.KSA}] \tag{12}$$

$$\Rightarrow k_b[\text{ATP}]\Big([\text{KSA}] - [\text{ATP.KSA}]\Big) = (k_h + k_-)[\text{ATP.KSA}] \tag{13}$$

so that

$$[\text{ATP.KSA}] = \frac{[\text{ATP}][\text{KSA}]}{\frac{k_h+k_-}{k_b} + [\text{ATP}]} \tag{14}$$

Extensile activity results from the relative sliding of microtubules past one another. We therefore argue that is should depend linearly on the number of steps exerted by the ATP-bound motor clusters walking on the microtubules, hence, the concentration of ATP bound motor clusters [ATP.KSA]:

$$\zeta = \zeta_0[\text{ATP.KSA}] \tag{15}$$

where $\zeta_0$ is a proportionality constant that depends on the efficiency with which ATP hydrolysis translates to mechanical motion.

## 2.2 Shear and bend moduli

There are two kinds of cross-linkers that are present in the system:

1. Non-motile kinesin motors: The concentration of these are given by $[\text{KSA}]_0$=[KSA]−[ATP.KSA]
2. PRC1 cross-linkers: As we did for KSA, we assume that there is an excess of tubulin compared to PRC1 so that almost all PRC1 used to prepare the solution can be assumed to be bound to microtubules.

Thus, the total concentration of cross-linkers is given by [cross-linker]=[KSA]−[ATP.KSA]+p0.[PRC1]. PRC1 being a sensitive protein to purify, we considered that only a fraction p0 of the proteins are active. These cross-linkers generate elasticity in the network by constraining the motion of parts of the network relative to each other.

How the mechanical properties of purified cytoskeletal networks vary with the concentration of crosslinking proteins has been studied for a number of systems. Above a critical concentration of crosslinking proteins, the elastic plateau modulus has been found to scale with the concentration of crosslinkers, c, as $G_0 \sim c^\gamma$. Reported values for $\gamma$ vary for different crosslinkers, and frequently fall within the range $0.4 \leq \gamma \leq 2$ [8, 9, 10, 11, 12]. While such scalings have been measured for a number of different actin crosslinking proteins, microtubule networks have been less explored. As microtubules are much more rigid than actin filaments, we here use the result from percolation theory: the elastic moduli should scale quadratically with the cross linker concentration at low cross linker concentrations[13, 8, 14]. Therefore, we postulate that the shear modulus and the bend modulus should be of the form

$$\mu = \mu_0[\text{cross-linker}]^2 \tag{16}$$

$$\kappa = \kappa_0[\text{cross-linker}]^2 \tag{17}$$

## 2.3 Nematic elasticity

Since the instability in the experiment consist of bend deformations (gradients in the deformation are along the initial direction of order, $\hat{x}$), it is appropriate to assume that $K$ scales with the length $l$ of microtubules as $K \sim l^3$, as is appropriate for the bend constant of liquid crystals [15].

$$K = K_0 l^3 \tag{18}$$

## 2.4 Fitting parameters

Using the expressions for $\mu$, $\kappa$, $K$, and $\zeta$ in the equation for phase boundary,

$$\left(\frac{\zeta_0}{\mu_0}\frac{[\text{ATP.KSA}]}{[\text{cross-linker}]^2}\right)\left(1-\frac{1}{\sqrt{1+\frac{\kappa_0}{K_0}\frac{[\text{cross-linker}]^2}{l^3}}}\right)=1 \tag{19}$$

Similarly, the out of plane wavelength is given by

$$\lambda_{\text{OP}} = 2\sqrt{2}\pi\sqrt{\frac{K_0}{\zeta_0}}\sqrt{\frac{l^3}{[\text{ATP.KSA}]}\left(1+\frac{\kappa_0}{K_0}\frac{[\text{cross-linker}]^2}{l^3}\right)} \tag{20}$$

Therefore, the fitting parameters are the three ratios, $r_1 = \frac{\zeta_0}{\mu_0}$, $r_2 = \frac{K_0}{\kappa_0}$, $r_3 = \sqrt{\frac{K_0}{\zeta_0}}$, and p0, the ratio of active PRC1 proteins.

# 3 Parameter Estimation through Markov Chain Monte Carlo

## 3.1 Fitting algorithm

We assume that all data points that have a blurriness value $B < B_0$ correspond to in plane instability, and those with $B > B_0$ correspond to out of plane instability (we chose $B_0 = 0.1$ based on experimental observations). In order to fit the experimental data to the theoretical model, we define a phase parameter $P_e = \Theta(B - B_0)$, where $\Theta$ is the Heaviside step function. $P_e$ is unity when the instability is out of plane, and vanishes when the instability is in plane. The corresponding theoretical prediction for the phase parameter is given by $P_t = \Theta(\frac{\zeta}{\mu} - \frac{1}{1-\frac{1}{\sqrt{1+\frac{\kappa}{K}}}})$. For a given data point in the phase diagram, and a proposed set of parameters $r_1, r_2$, the quantity $|P_e - P_t|$ vanishes if the theoretical prediction is correct, and is equal to unity otherwise. The number of data points that the theory correctly predicts is $n = \sum_{i=1}^{N} |P_e(i) - P_t(i)|$, where $N$ is the total number of data points from all four phase diagrams. Suppose we expect the theoretical model to predict each data point correctly with probability $p$ ( $p$ is a measure of the uncertainty in the experimental phase boundary. We chose $p = 0.99$). The likelihood that a given set of parameters correctly predicts all four experimental phase diagrams $\sim p^n(1-p)^{N-n}$. Further, we assume that the probability that a set of parameters $r_2, r_3$ correctly predicts the experimental wavelengths $\sim \Pi_{j=1}^{M} e^{-\frac{(\lambda_e(j)-\lambda_t(j))^2}{(4\sigma_\lambda^2)}}$, where $\lambda_e$ is the experimentally measured wavelength, $\lambda_t$ is the corresponding theoretical wavelength predicted by the parameter set, $\sigma_\lambda$ is the standard deviation of the experimental wavelength measurement, which we took to be $50\mu m$, and $M$ is the number of data points for wavelength. Thus, the total likelihood that a set of parameters $r_1, r_2, r_3$ correctly predicts experimental data $\sim p^n(1-p)^{N-n}\Pi_{j=1}^{M} e^{-\frac{(\lambda_e(j)-\lambda_t(j))^2}{(4\sigma_\lambda^2)}}$.

The parameters were estimated by performing Markov chain Monte Carlo moves in the estimated parameter range. During each iteration, a new set of parameters were proposed by picking parameter values from a uniform distribution over the estimated range. If the likelihood associated with the proposed set of parameters is higher than that for the old set of parameters, the parameters were updated with the newly proposed values. If the likelihood associated with the proposed set of parameters is lower than that for the old set of parameters, a uniform random number between 0 and 1 was generated, and the new parameters were accepted only if the ratio of likelihood associated with the new parameters to that of the old parameters was larger than the random number. We started our simulations both from the upper and the lower limits of the estimated parameter range. In all cases, the Markov chain converged to the set of parameters that we report in Fig. S11. We chose the set of parameters with the maximum likelihood:

$$
\begin{aligned}
r1 &= 4810nM \\
r2 &= 7.02nM^2.\mu m^{-3} \\
r3 &= 0.602\mu m^{-1/2}.nM^{1/2} \\
p0 &= 0.592
\end{aligned}
$$

We therefore estimate that only 59% of the PRC1 is actually crosslinking the microtubules.

## 3.2 Range of fit parameters

In order to specify a range from which parameters are picked in the Markov Chain Monte Carlo method, we need to estimate the plausible range of parameter values, given the theoretical phase boundary and the range of experimental concentrations.

### 3.2.1 $r_1$

The in-plane instability exists only for $\zeta > \mu$. Therefore, we can assume $r_1 = \frac{\zeta_0}{\mu_0} > \frac{[\text{cross-linker}]^2}{[\text{ATP.KSA}]}$ while computing the phase boundary. Since KSA concentration in the experiments ranges between $1 - 200nM$, PRC1 concentration ranges between $60 - 380nM$, and ATP concentration ranges between $1\mu M - 1mM$, [cross-linker] ranges between $1 - 580nM$, and [ATP.KSA] ranges between $0.02 - 190nM$. Therefore, $\frac{[\text{cross-linker}]^2}{[\text{ATP.KSA}]}$ ranges between $10^{-3}nM - 17mM$.

### 3.2.2 $r_2$

The nematic elasticity for microtubules of length $1.5\mu m$ in 3D is around $1.6 \times 10^{-9}N$ [16], and the shear modulus for microtubule networks in 3D is around $60Pa$[17]. Since the thickness of the sheet is around $80\mu m$, the 2D nematic elasticity, $K \sim 1.6 \times 10^{-9}N \times (80\mu m)$, and the 2D bending modulus, $\kappa \sim 60Pa \times (80\mu m)^3$. Thus, $r_2 = \frac{K_0}{\kappa_0} = \frac{K/l^3}{\kappa/([\text{cross-linker}]^2)} \sim \frac{1.6\times10^{-9}N/(1.5\mu m)^3}{60Pa\times(80\mu m)^2/([\text{cross-linker}]^2)}$, ranges between $10^{-3} - 415nM^2\mu m^{-3}$. Therefore, we give $r_2$ a range of $10^{-4} - 10^3 nM^2\mu m^{-3}$ for the fitting.

### 3.2.3 $r_3$

$r_3 = \sqrt{\frac{K_0}{\zeta_0}} = \sqrt{\frac{K}{\zeta}}\sqrt{\frac{[\text{ATP.KSA}]}{l^3}}$; $\sqrt{\frac{K}{\zeta}}$ has units of length, and should range between the minimum (microtubule length $1.5\mu m$) and maximum (max channel length $\sim 3cm$) length scales in the system. Therefore, $r_3$ ranges between $1.5\mu m \times \sqrt{\frac{\min([\text{ATP.KSA}])}{\max(l)^3}} \sim 0.022nM^{1/2}\mu m^{-1/2}$ to $3 \times 10^4\mu m \times \sqrt{\frac{\max([\text{ATP.KSA}])}{\min(l)^3}} \sim 5.8 \times 10^5 nM^{1/2}\mu m^{-1/2}$

### 3.2.4 $p_0$

The minimum value of $p_0$ is zero, which corresponds to all PRC1 crosslinkers being inactive, and the maximum value of $p_0$ is 1, which corresponds to all PRC1 crosslinkers being actively crosslinking.

## 3.3 Estimation of $\mu_0$

Given the parameters $r_1$, $r_2$, and $r_3$, prior knowledge of $\mu_0$ allows us to infer the values of $\zeta_0$, $\kappa_0$ and $K_0$. Recent bulk rheology experiments estimated the storage modulus G'=60Pa for a microtubule-kinesin network that ran out of ATP. In that case, the concentration of crosslinkers is equal to the initial concentration of molecular motors 120nM. We therefore estimate that $\mu_0 = 60Pa/(120nM)^2 = 4.17 \times 10^{15} Pa.M^{-2}$

## 3.4 Comparison of the value of $\zeta$ with values from the literature

Activity has been inferred previously by Ellis and co-workers in [16]. Briefly, they inferred the activity from tracking topological defects in 2D active nematics composed of microtubules and molecular motors. The value they measured is around 250mPa, which is about 3 orders of magnitude lower than the values reported here for the same concentrations of motors and ATP. There are a few differences between the two experiments that could explain this difference. First, Ellis et al. studied a 2D active nematic on an oil-water interface. Here, we have a suspension of microtubule bundles in 3D. Second, the local microtubule density is much larger in their study as they deplete all the microtubules on an oil-water interface. Lastly, we note that the activity has to be larger than the elasticity for any liquid like instability, which for the protein concentrations reported by Ellis [16] and by Gagnon in [17] mean $\zeta > \mu = 60Pa$

# 4 Materials and Methods

## 4.1 Protein purification protocols

### 4.1.1 Microtubules (MTs)

Tubulin dimers were purified from bovine brains through two cycles of polymerization - depolymerization in high molarity PIPES (1,4- piperazindiethanesulfonic acid) buffer [18]. Fluorophore-labeled tubulin was prepared by labeling the purified tubulin with Alexa-Fluor 647-NHS (Invitrogen, A-20006)[19]. GMPCPP (Guanosine 5'- ($\alpha$, $\beta$ methylenetriphosphate)), a non-hydrolyzable analogue of GTP was used to stabilize the dynamic instability in the MTs. Polymerization mixture consisted of 80$\mu$M tubulin (with 3% fluorescently labeled tubulin), $0.6mM$ GMPCPP and $1mM$ DTT (dithiothreitol) in M2B buffer ($80mM$ PIPES, $1mM$ EGTA, $2mM$ $MgCl_2$). After adding all the components, the mixture was incubated at 37°C for 30 minutes, and subsequently for 6 hours at room temperature ($\sim$ 20°C). This method resulted in Microtubules of $\sim$ 1.5$\mu$m length [20]. The stock concentration was 8mg/mL. Microtubules were aliquoted in small volumes (10$\mu$L), flash-frozen in liquid nitrogen, and stored at -80°C.

### 4.1.2 Crosslinkers - PRC1-NSΔC

The truncated PRC1-NSΔC (MW : 58 kDa), a microtubule crosslinking protein, was expressed and purified in Rosetta BL21(DE3) cells using an established protocol described elsewhere [21]. The proteins were flash frozen with 40% sucrose and stored at -80°C. The final concentration of PRC1 was measured by a Bradford assay.

### 4.1.3 Molecular motors - K401 and K365 motors

K401-BIO-6xHIS (processive motor, dimeric MW-110 kDa) and K365-BIO-6xHIS (non-processive motor, MW-50 kDa) are biotinylated kinesin constructs derived from N-terminal domain of *Drosophila melanogaster* kinesin-1, truncated at residue 401 and 365, respectively, and labeled with six histidine tags. The motor proteins were transformed and expressed in Rosetta (DE3) pLysS cells and purified following established protocols described previously [22]. The proteins were stored in a 40% wt/vol sucrose solution at -80°C. The final concentration of kinesin was measured by Bradford assay.

We used tetrameric streptavidin (ThermoFisher, 21122, MW: 52.8 kDa) to assemble clusters of biotinlabeled kinesins (KSA). To make $K401$-streptavidin clusters, 5.7$\mu$L of 6.6$\mu$M streptavidin was mixed with 5$\mu$L of 6.4$\mu$M $K401$ and 0.5$\mu$L of 5mM DTT in M2B. This mixture was incubated on ice for 30 minutes. K365-streptavidin clusters were prepared by mixing, 5.7$\mu$L of 6.6$\mu$M streptavidin, 3.1$\mu$L of 20$\mu$M K365, 0.5$\mu$L of 5$mM$ DTT and 1.94$\mu$L of M2B, and then left to incubate on ice for 30 minutes

#### 4.1.4 Light activable motors: $K365$-iLID and $K365$-micro

K365-iLID and K365-micro were designed by Linnea Lemma and Tyler Ross and were purified following the protocol described in [23]. Briefly, two chimeras of the *Drosophila melanogaster* kinesin K-365, namely, K-365-iLID and K-365-micro, were expressed in Escherichia coli Rosetta 2(DE3)pLysS cells and purified using ÄKTA pure FPLC system. MBP domain was cleaved with TEV protease (Sigma Aldrich). The proteins were snap-frozen in 40% glycerol and stored at -80°C.

### 4.2 Assembling 3D active network

The networks are composed of:

- Alexa 647-labeled GMPCPP stabilized microtubules (MTs) with an exponential distribution of lengths with an average of 1.5$\mu$m, unless otherwise specified,
- Multi-motor kinesin complexes self-assembled from tetrameric streptavidin and two-headed biotinylated kinesin ($K401$-Bio) or single-headed biotinylated kinesin ($K365$-Bio) [24],
- a specific microtubule bundling protein, PRC1 (the protein regulator of cytokinesis 1), that passively crosslink antiparallel microtubules, but still allow interfilament sliding [21].

In the presence of ATP, these proteins form a network of extensile bundles. An ATP regeneration system (phosphoenol pyruvate (26$mM$ PEP, Beantown Chemical, 129745) and pyruvate kinase/lactate dehydrogenase enzymes (2.8% v/v PK/LDH, Sigma, P-0294) was used to sustain a constant ATP concentration. An oxygen scavenging system comprised of glucose (18.7$mM$), DTT (5.5$mM$), glucose oxidase (1.4$\mu$M), and catalase (0.17$\mu$M) was used to decrease photobleaching. Active network composed of MTs (1.3mg/ml), ATP (1420$\mu$M) and K401 clusters (120$nM$) remain active for 6-8 hours. For some experiments (Fig. S3), PRC1 was replaced by either 20 kDa PEG (polyethylene glycol) (0.8% (wt/vol) [Sigma] or Pluronic F-127 2% (wt/vol) [F-127, Sigma $P2443$. MW: 12.5 kDa]. For consistency, a large volume of premix was prepared, aliquoted, and snap frozen in liquid nitrogen for each phase diagrams shown in Fig. 3. Frozen microtubules (stored at -80°C) were thawed immediately before use in an experiment. All the experiments were performed at room temperature.

### 4.3 Estimation of the ratio of motor clusters and crosslinkers per microtubules

For a microtubule of 1.5$\mu$m length, there are about 2444 tubulin dimers in its 13 protofilaments. If one tubulin dimer weighs 100kDa then one microtubule weighs $244,400$ kDa. Each active network contains 1.33mg/mL of microtubules which gives MT concentration of 5.45$nM$. Example: assuming all the clusters are bound to MTs, for [Motor cluster]= 120$nM$ and [PRC1]=60$nM$, they are about 22 motor clusters and 11 crosslinkers per microtubules.

### 4.4 Controlling the length distribution of microtubules via end-to-end annealing

We control the length distribution of GMPCPP-stabilized, fluorescently labeled microtubules by end-to-end annealing at 37°C for different time duration (Fig. S10). The polymerization protocol described above produces MTs with an average length of 1.5$\mu$m at a tubulin concentration of 8mg/mL. Below, we describe how to increase the average microtubule length at a constant tubulin concentration. The slowly hydrolysable GTP analog, GMPCPP effectively reduces the microtubule dissociation [25], therefore, the increase in length can only be attributed to end-to-end joining and thus other mechanisms including dynamic instability, subunit exchange at the MT ends or nucleotide hydrolysis can be ignored [26]. Keeping the MT in the water bath at 37°C increase the average MT length (Fig. S3-d).

## 4.5 Flow chamber assembly

All flow chambers have the same dimensions (H=80$\mu$m, W=3$mm$) and were assembled using two different approaches. No differences were observed between the 2 methods. The first method relies on a microfabricated channel made out of PDMS. The channel was prepared from a molding master created by machining cyclic olefin copolymer (COC) [27]. PDMS was poured over the master, and cured for 1 h at 70°C. The channel was removed from the master, inlet and outlet were punched, followed by oxygen plasma bonding to a glass slide (26 × 75 × 1$mm$). The channel was incubated with 2% Pluronic solution for 30 minutes to block any non-specific protein adsorption. The second method consists in two glass slides spaced by a layer of parafilm. The glass surfaces were coated with an acrylamide brush to resist non-specific protein adsorption onto the glass due to depletion [28]. Parafilm spacers were cut and placed between the two glass surfaces followed by a mild heat treatment at 60°C to melt parafilm so it can bind to the glass surfaces. The active mixture was loaded into each channel type by capillarity and sealed with an UV-curing optical adhesive (NOA 81, Norland Products Inc.).

## 4.6 Microscopy

### 4.6.1 Widefield microscopy

The Alexa 647 labeled microtubule networks were imaged using an inverted wide-field microscope (Nikon Ti-E or Ti2) with a fluorescent filter (Semrock Cy5-4040C), a SOLA light engine (Lumencor), a 10x objective (Nikon Pan Fluor, NA 0.3) and a CCD or a sCMOS camera (Andor Clara E, Hamamatsu orca flash 4.0). The illumination and the data acquisition were controlled by micro-manager ($\mu$Manager, Version 2.0.0-gamma [29]). All the measurements were performed at room temperature.

### 4.6.2 Confocal microscopy

Confocal fluorescence images of microtubules were obtained under Leica Application Suite X (LAS X) control on a laser scanning confocal microscope (TCS-SP8, Leica Microsystems GmbH) equipped with photomultiplier tubes. Fluorescence was excited with 638$nm$ laser diode for Alexa Fluor 647. For morphological analysis, a z-stack with a 775 × 775$\mu$m field of view was scanned with a step size of 2$\mu$m using a non-immersion 20x objective (HCX PL Fluotar, numerical aperture, NA = 0.50).

## 4.7 Light-activable motors: microscopy Protocol and light intensity measurement

We replace Kinesin-streptavidin clusters by light activable hetero-dimers K365-iLID and K365-micro. Active mixtures of microtubules were prepared with equimolar concentration (0.2$\mu$M) of K365-iLID and K365-micro, 30$\mu$M ATP, and the standard ATP regenerating system. The microtubule networks were imaged and photo-activated with an inverted Nikon Ti2 microscope equipped with 10X objective (Nikon Plan Apo $\lambda$ 10X, NA 0.45), a SOLA light engine (Lumencor), and a sCMOS camera (Andor Zyla). The illumination and the data acquisition were controlled by micro-manager. All the measurements were performed at room temperature. Light activable motors were excited at 488$nm$ and fluorescent images of Alexa647-labeled microtubules were acquired using a Cy5 filter. Only a predefined region of interest (ROI) of total field of view of the camera was illuminated. Typically, one experiment was run per sample. In all experiments, the input power, $P_{\text{input}}$ of the blue light was measured with an optical power meter (PM100A, Thorlabs GmbH, Germany) at the sample plane at $488nm$. The total output power ($P_{\text{total}}$) on the sample surface from a blue light beam was calculated by:

$$P_{\text{total}} = P_{\text{input}} \cdot \frac{\text{Exposure time (s)}}{\text{time interval (s)}} \tag{21}$$

The illumination intensity (I) was calculated by:

$$I = \frac{P_{\text{total}}}{A} \tag{22}$$

where A is the area of the sample. Samples were prepared and flown into the microfluidic chambers in the dark.

## 4.8 Image Processing

### 4.8.1 Blur segmentation algorithm

The microtubule bundles are blurry when they buckle outside of the focal plane of the microscope. We developed an image analysis protocol to detect and segment such out-of-focus patches (Fig. S2). We measured the magnitude of the gradient of the fluorescent intensity – aka the image sharpness. When the microtubules are out of focus, the sharpness is low. When they are in focus, the sharpness is large. The out-of-focus patches are usually large (characteristic size: 50-100$\mu$m), we therefore smoothed the sharpness to remove high frequency noise. Finally, we segmented the sharpness to identified the out-of-focus regions. All the pixels with a sharpness below a threshold value are considered out of focus, while all the others are in-focus. Overlay between the segmented out-of-focus patches and the original fluorescent images confirmed the effectiveness of this detection method. All the images were acquired with the same objective, exposure time, filter cube, light intensity and binning.

$$B = \frac{\text{Area of the out-of-focus patches}}{\text{Total area of the Field Of View}} \tag{23}$$

### 4.8.2 In- plane flow measurement

In-plane flows (Fig. S1) were measured during the instability growth using MATLAB-based open-source particle image velocimetry (PIV) [30]. An interrogation window size of 64 pixels = 82.56$\mu$m was selected with 50% overlap.

### 4.8.3 Measurement of the in-plane instability wavelength

The wavelength of the in-plane instability was measured from heatmaps of the y-component of the velocity field using a custom-made Fast Fourier Transform (FFT) algorithm written in MATLAB (Fig. S8). FFT of the velocity profiles was computed along the initial axis of alignment. The resulting power spectrum revealed a strong periodic signal with a well-defined peak corresponding to a characteristic wavelength of in-plane instability.

### 4.8.4 Measurement of the out of plane instability wavelength

The wavelength of the instability was measured from a Fast Fourier Transform of the sharpness (Fig. S9, see section above about how to measure the sharpness). The wavelength of the out-of-plane buckling instability is equal to twice the periodicity of the sharpness as it does not differentiate between microtubule bundles above or below the focal plane.

#### 4.8.5 Measurement of the microtubules' length distribution

To characterize the length distribution of the microtubules, we sampled the MT stock at a constant tubulin concentration (8mg/mL) after every 30 minutes over 5 hours of annealing at 37°C. The samples were diluted to 3000X with antioxidants, Trolox, and a 2.5% (w/v) solution of dextran (MW 500 kDa). The dilution helps to prevent overlaps of MTs that complicate automated filament recognition. For imaging fluorescently labelled microtubules, 5$\mu$L of the solution was placed between a coverslip and a coverslide. Imaging was performed on a standard fluorescence microscope (Nikon Eclipse Ti microscope) using a high numerical aperture oil objective (Nikon Plan Flour 100X/1.30). We wrote a custom Matlab script that quantifies the length distribution of MTs. Binary images of each MT were fitted with an ellipse. The long axis of the ellipse corresponds to the MT's length (Fig. S10).

# 5 Supplementary Figures

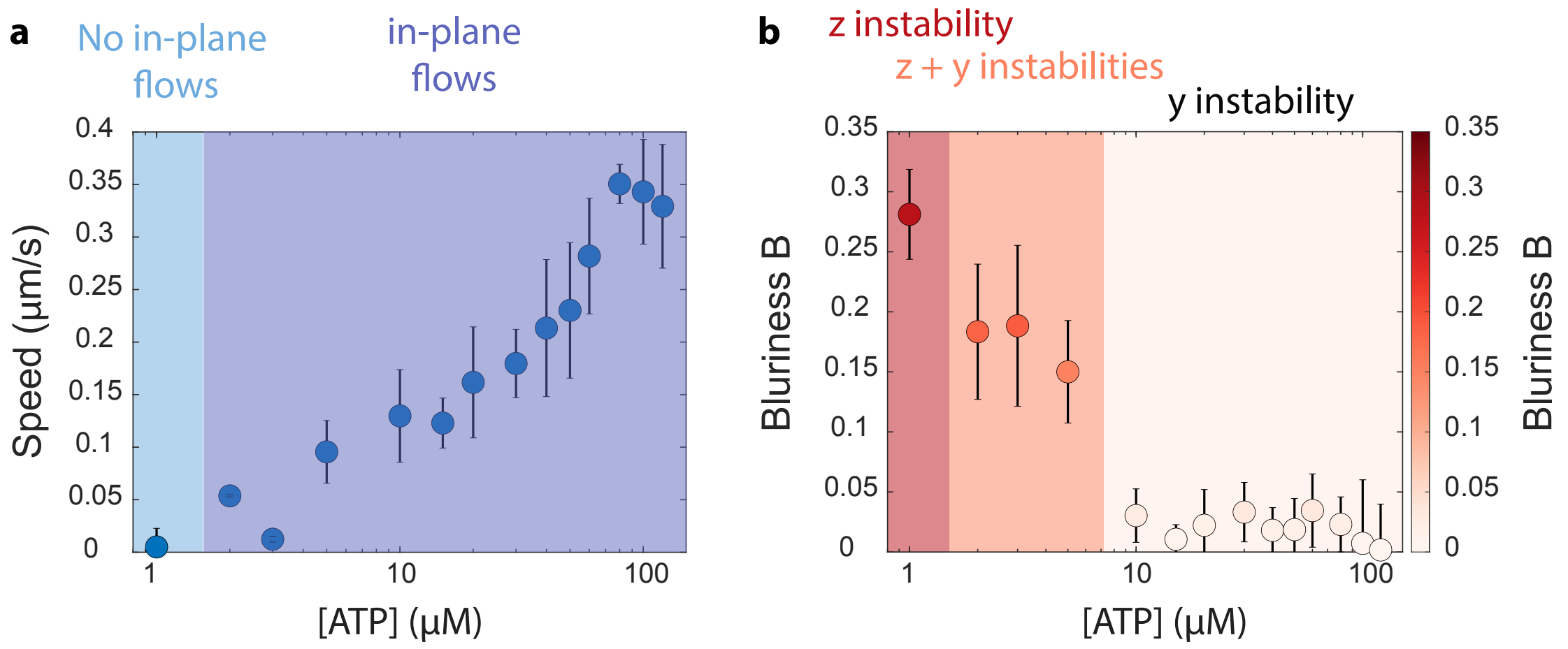

Fig. S1: **In-plane flows are suppressed at low ATP concentration**. a) in-plane flows measured by Particle Image Velocimetry and b) blurriness as a function of ATP concentration. No in-plane flows were observed for [ATP] $< 2\mu$M. Above that, the instability is a superposition of in-plane and out of plane deformations (2-5$\mu$M). Above 10$\mu$M, no deformations along the Z axis were detected. [motor clusters]$= 5nM$, [PRC1]$= 100nM$.

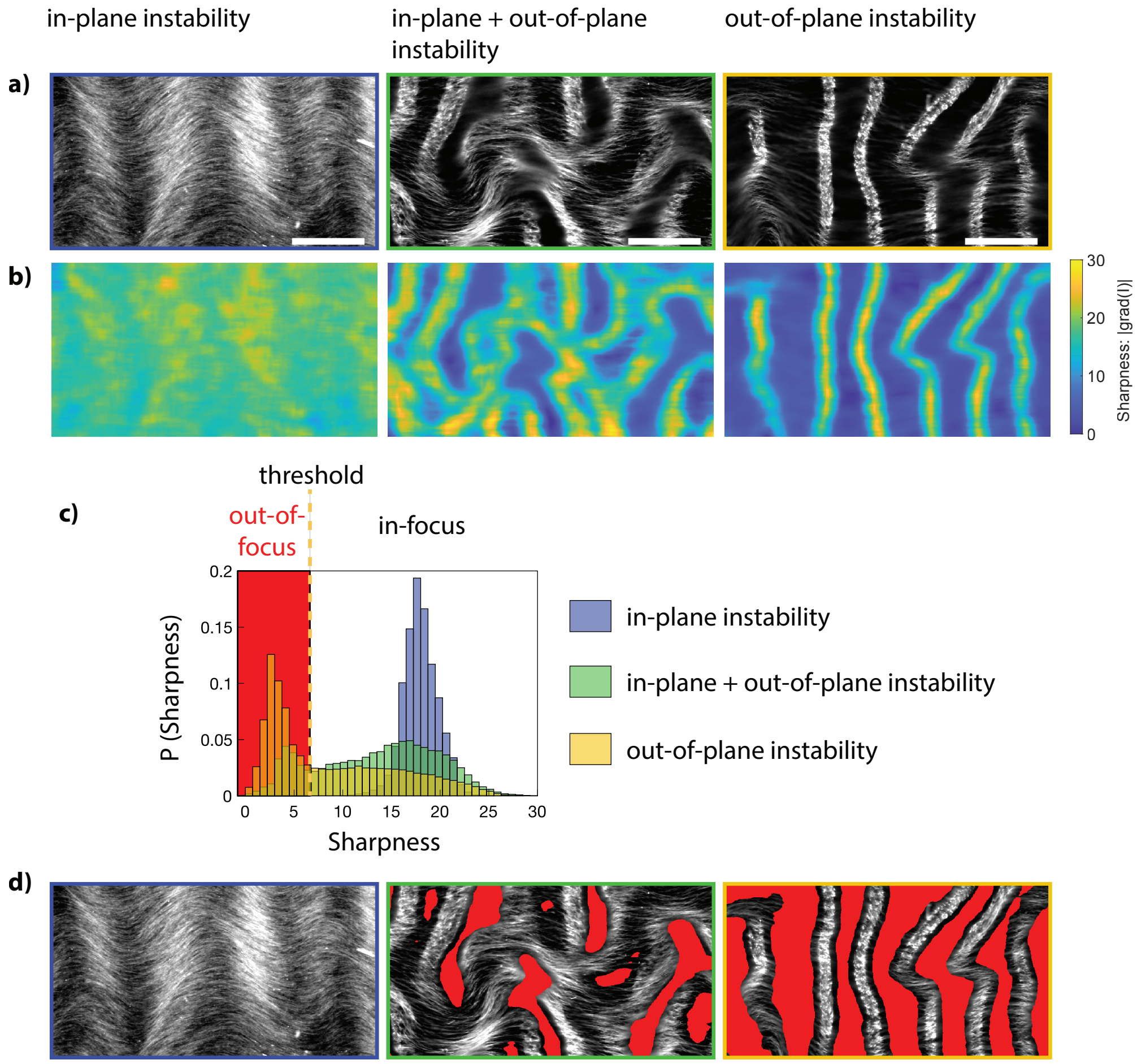

Fig. S2: **Method to detect if the network is bending in-plane or buckling out-of-plane.** a) Fluorescence images of the microtubule network (widefield fluorescent microscope, scale bar: 200$\mu$m). Three characteristic pictures are displayed showing an in-plane instability, an out-of-plane instability, or a superposition of both. b) heatmaps of the sharpness (magnitude of the gradient of the fluorescent intensity). High values of sharpness are associated with in-focus features, while low values of sharpness are associated with out-of-focus features. c) probability distribution of the sharpness for the three heatmaps in b). Pixels with sharpness valued below a fixed threshold are considered out-of-focus. d) Original fluorescent images onto which the segmented out-of-focus patches have been highlighted in red.

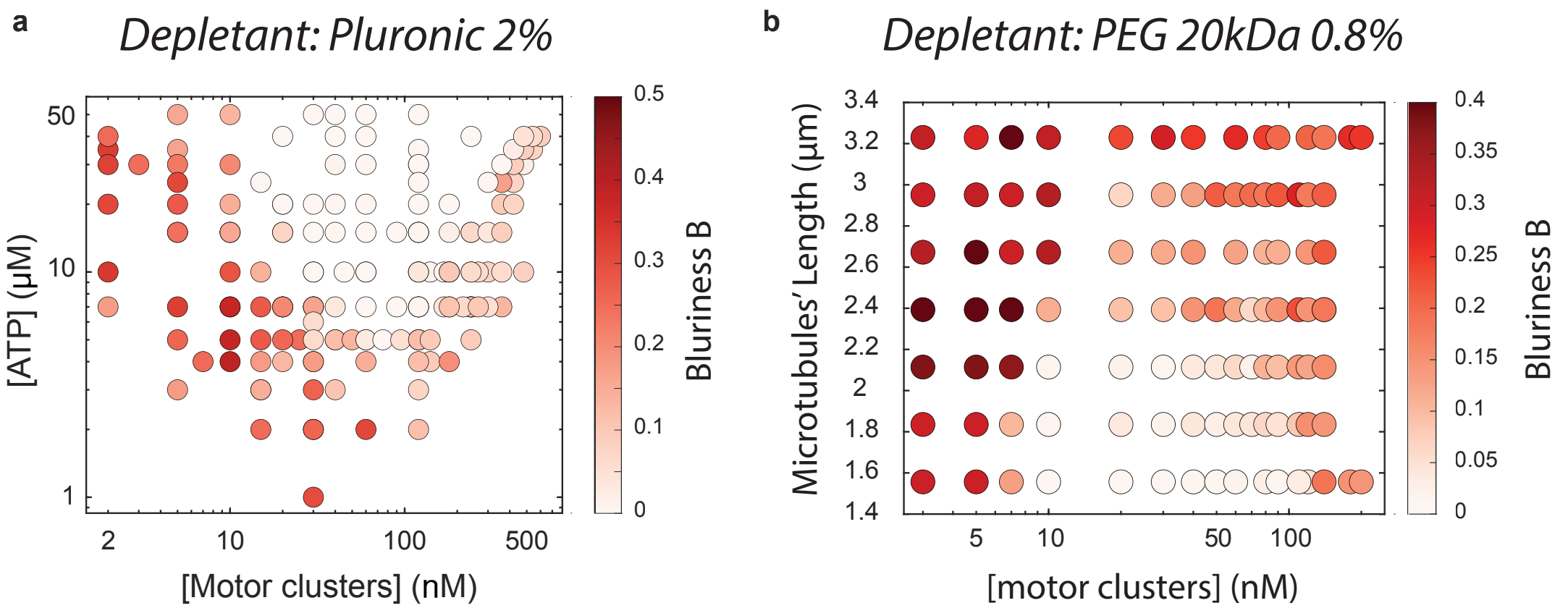

Fig. S3: **Increasing motor clusters concentration leads to a reentrant transition even when the crosslinker PRC1 is replaced by a non-specific bundler (2% Pluronic or 0.8% 20kDa PEG)**. a) ATP vs motor cluster concentrations phase diagrams show the re-entrant transition. b) Microtubules' length vs motor cluster concentrations. Color bar shows blurriness

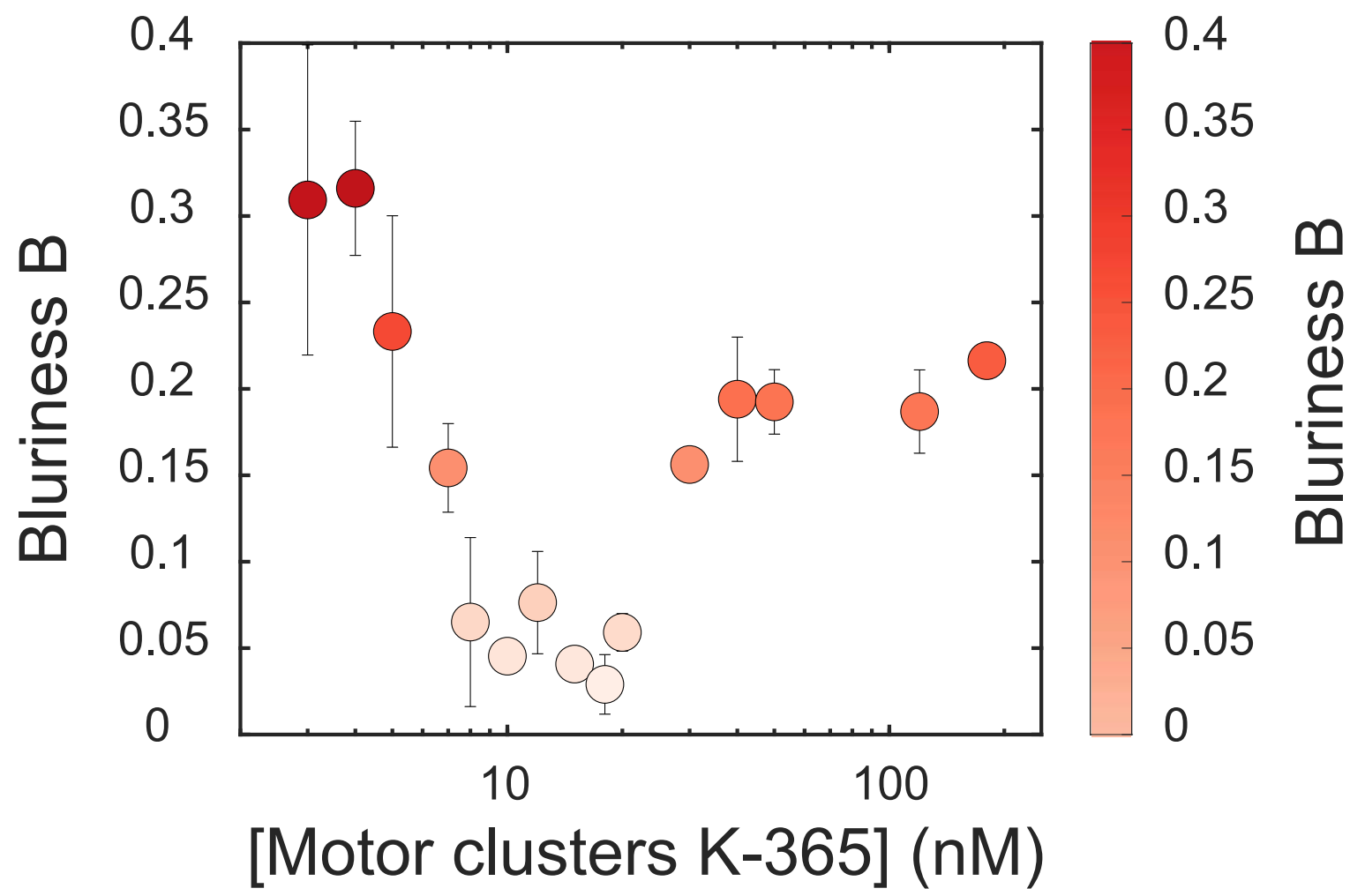

Fig. S4: **Non-processive K365 Kinesin motor clusters display the same re-entrant transition as the processive K401 motor clusters**. Color bar shows blurriness. [ATP]= 8$\mu$M, [PRC1]= $100nM$

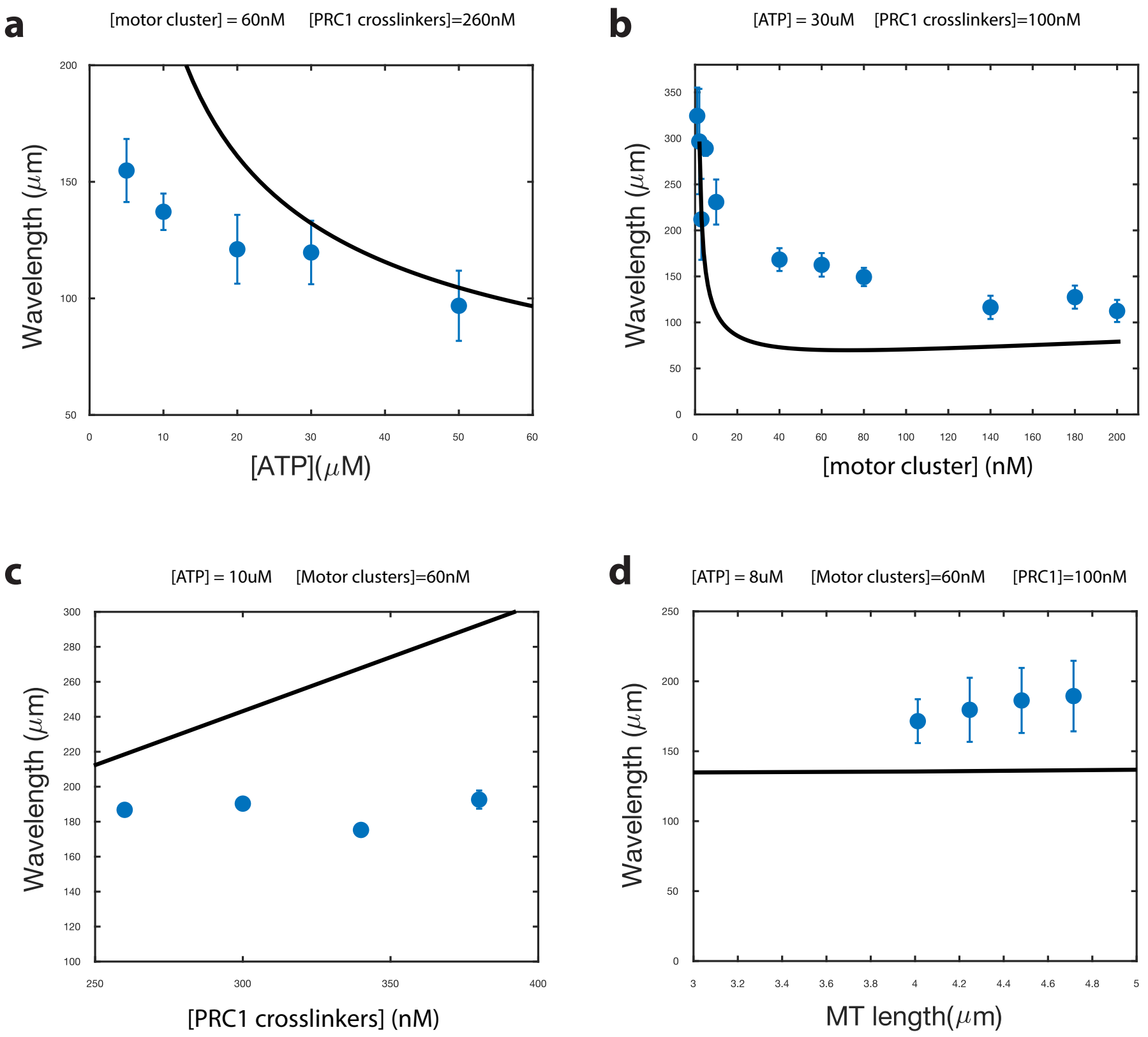

Fig. S5: **Comparison between theory and experiments for the out-of-plane wavelengths for various network compositions**. Blue points: experiments, Black lines: theory.

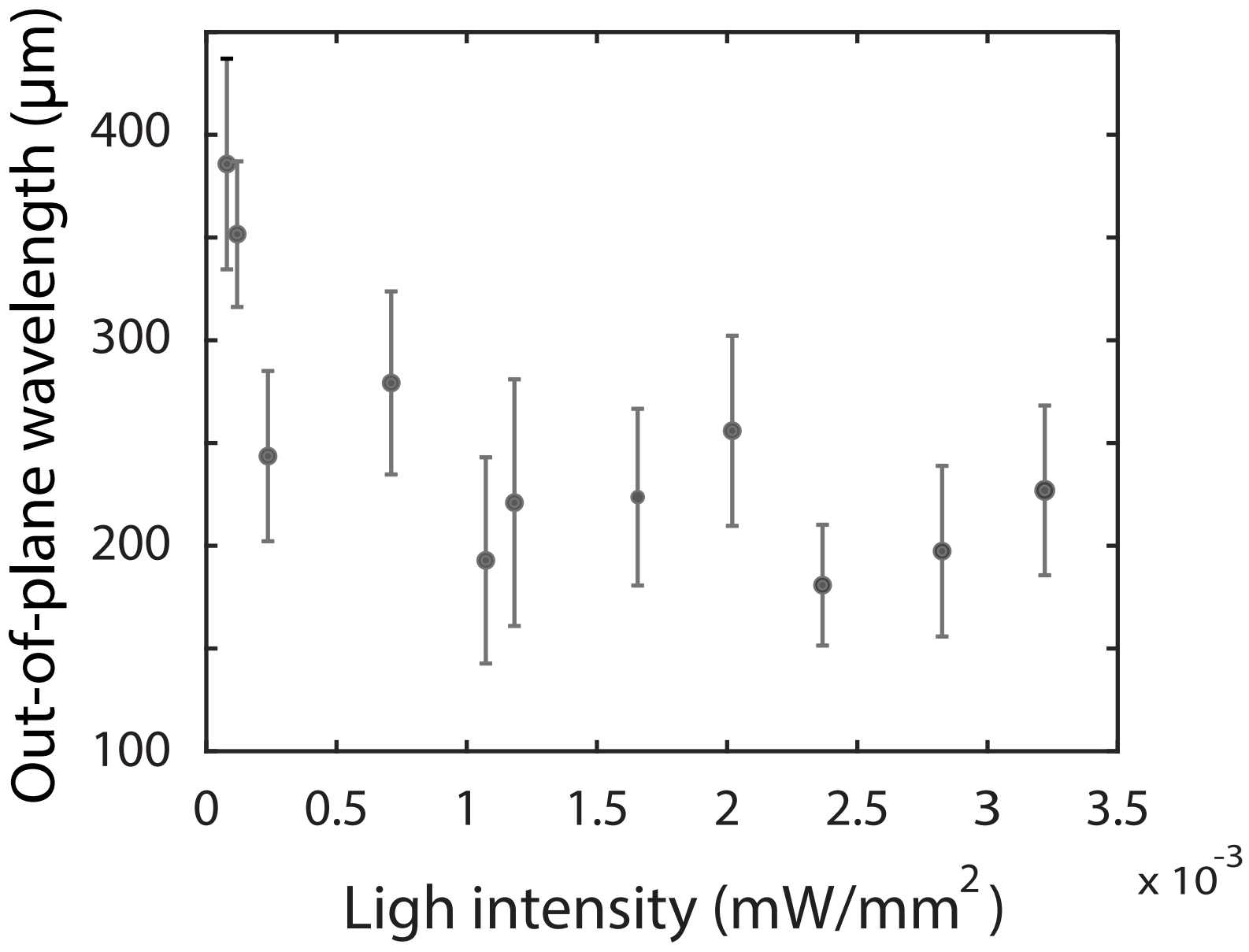

Fig. S6: **Out-of-plane wavelength for network composed of light-activable motors**. The wavelength decreases when the blue light intensity increases, which is compatible with an increase in activity. [ATP]=30uM, [LAMPS motors]=0.2uM.

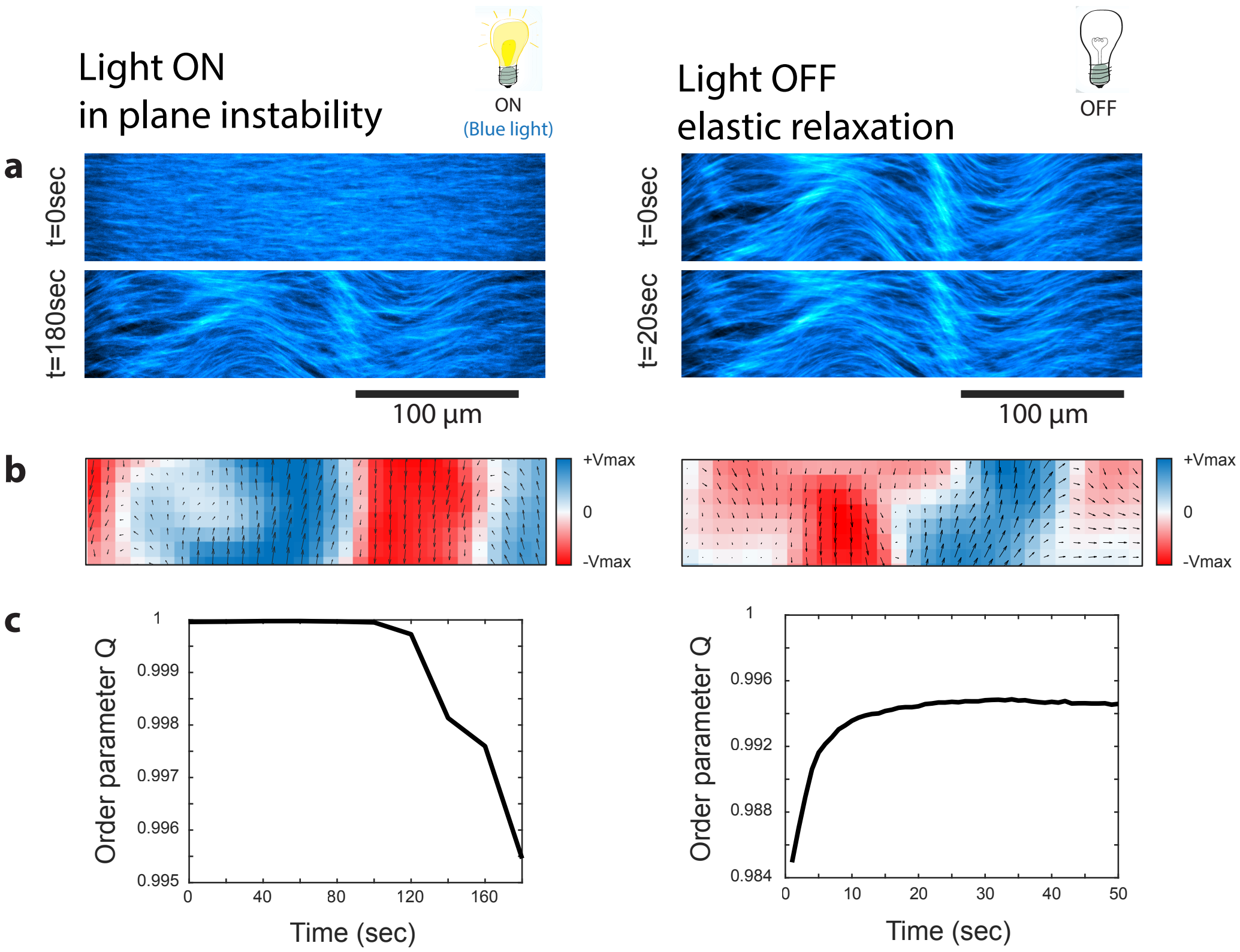

Fig. S7: **Turning off motor activity only allow a partial relaxation of the deformations**. Left column: active instability when blue light is one. Right column: elastic relaxation once the blue light is turned off. a) fluorescent image of the MT bundles. b) velocity field overlaid onto the heatmap of the y-component of the velocity. C) time evolution of the order parameter that quantifies the nematic alignment of the MT bundles. When the light is on, the instability grows and Q decreases. When the light is off, the deformations partially relax, Q increases but do not reach Q=1.

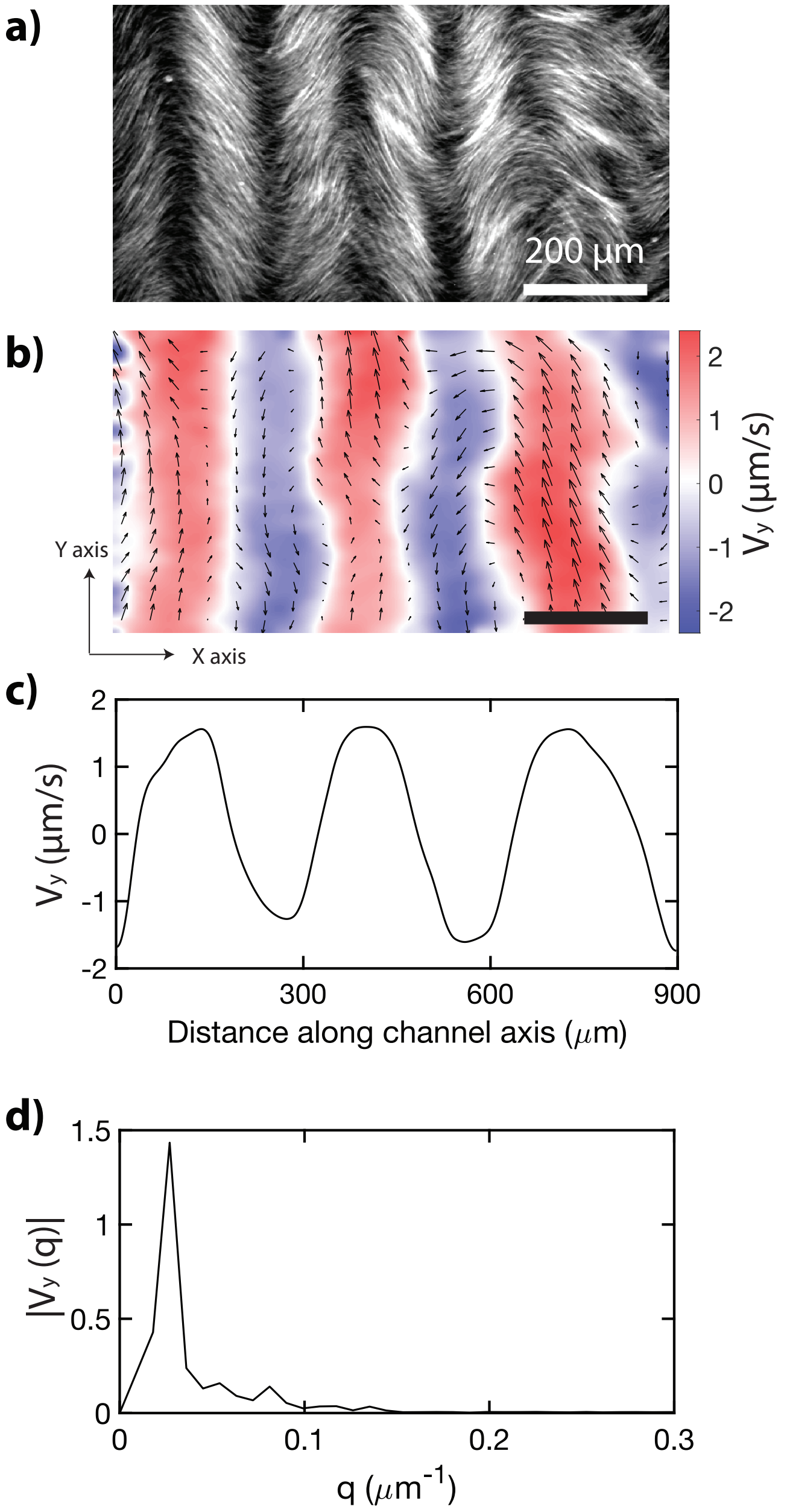

Fig. S8: **Characterization of the in-plane instability wavelength**. a) Fluorescent image of the microtubule bundles during the growth of the in-plane instability. b) velocity field (black arrows) overlaid onto the heatmap of the y-component of velocity field. c) Y-component of the velocity profile at fixed value of y, along the channel direction (x-axis). d) FFT spectrum of the corresponding velocity profile, q is the spatial frequency obtained by spatial domain of channel axis. $\lambda$ is estimated by calculating the characteristic periodic length from the spectrum.

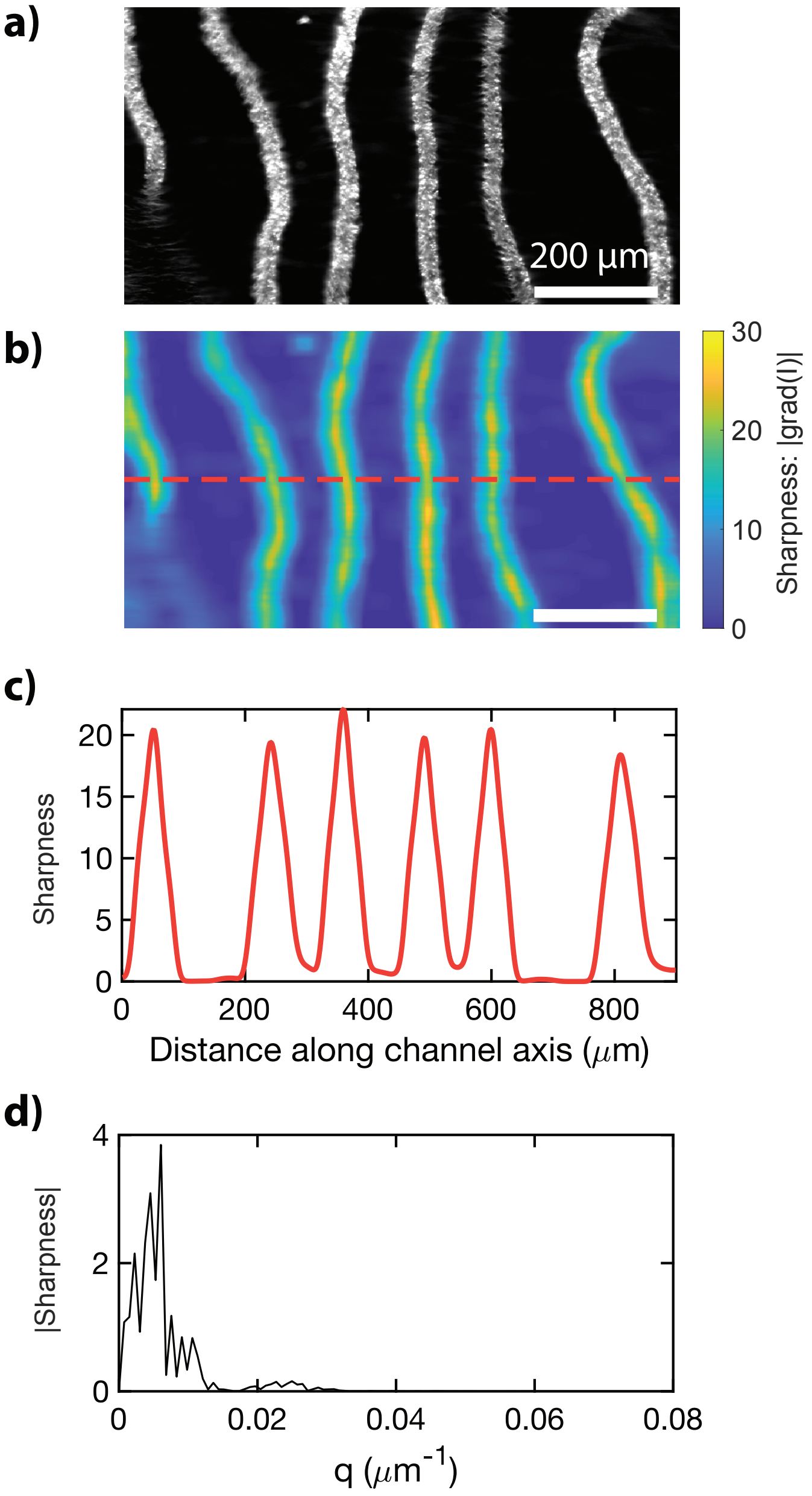

Fig. S9: **Characterization of the out-of-plane buckling wavelength**. a) fluorescent image of the microtubule bundles during the growth of the out-of-plane instability. b) Heatmap of the corresponding sharpness. c) A one-dimensional profile of the sharpness (at fixed value of y, red line shown in (b)) is used to calculate d) the power spectrum of the FFT.

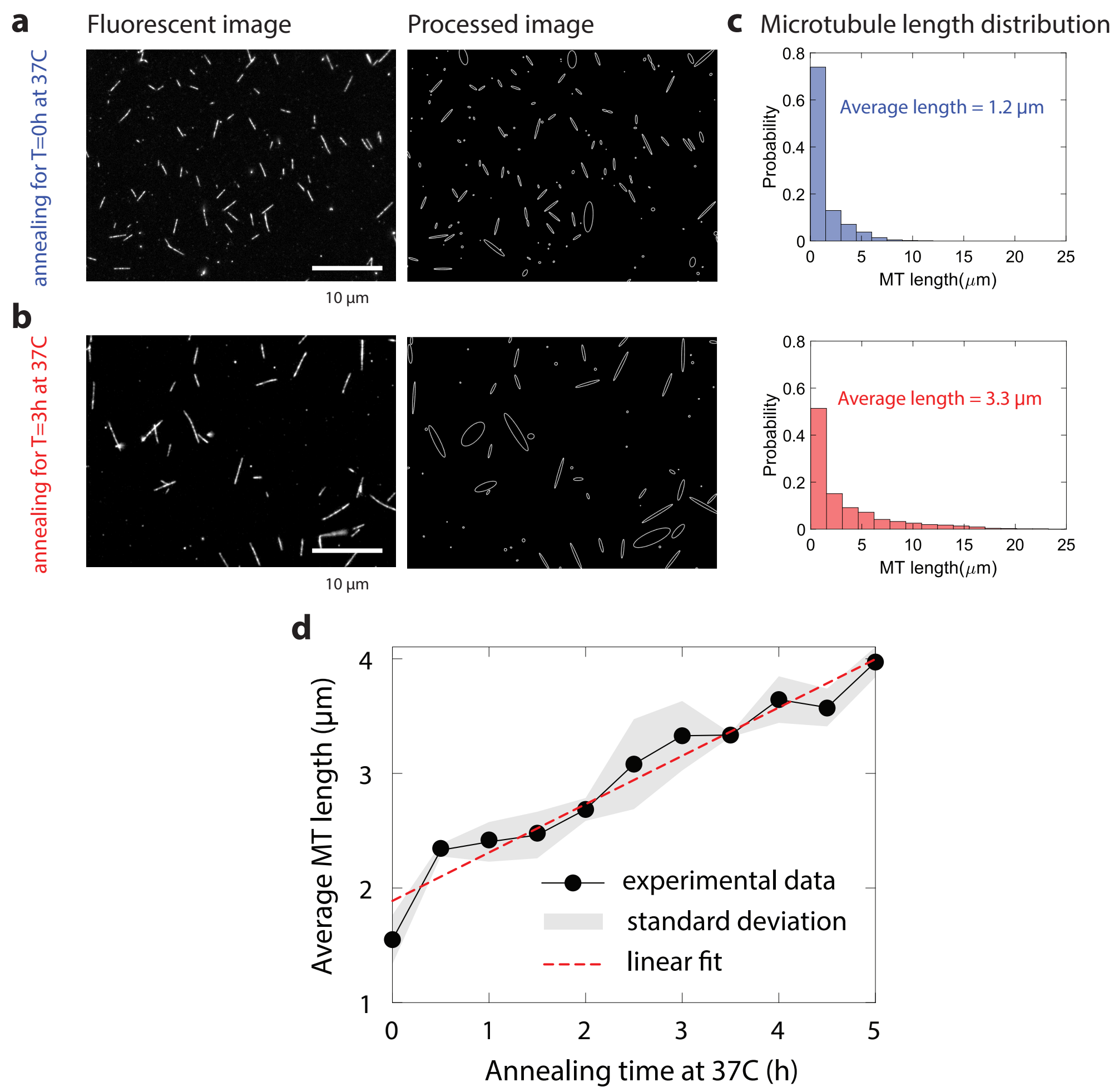

Fig. S10: **Controlling the length distribution of microtubules via end-to-end annealing**. a) Fluorescent images of the microtubules pre and post-annealing at 37°C. b) Ellipses are fitted onto the threshold fluorescent image to estimate the length distributions of the microtubules. c) Probability distributing of the microtubules' lengths before and after annealing for 3h at 37°C. The insert shows the averaged microtubule length. d) MTs' length as a function of annealing time at 37°C. The black dotted line is the experimental data and the grey area is the standard deviation in the experimental data over 3 realizations. The red dashed line is a linear fit.

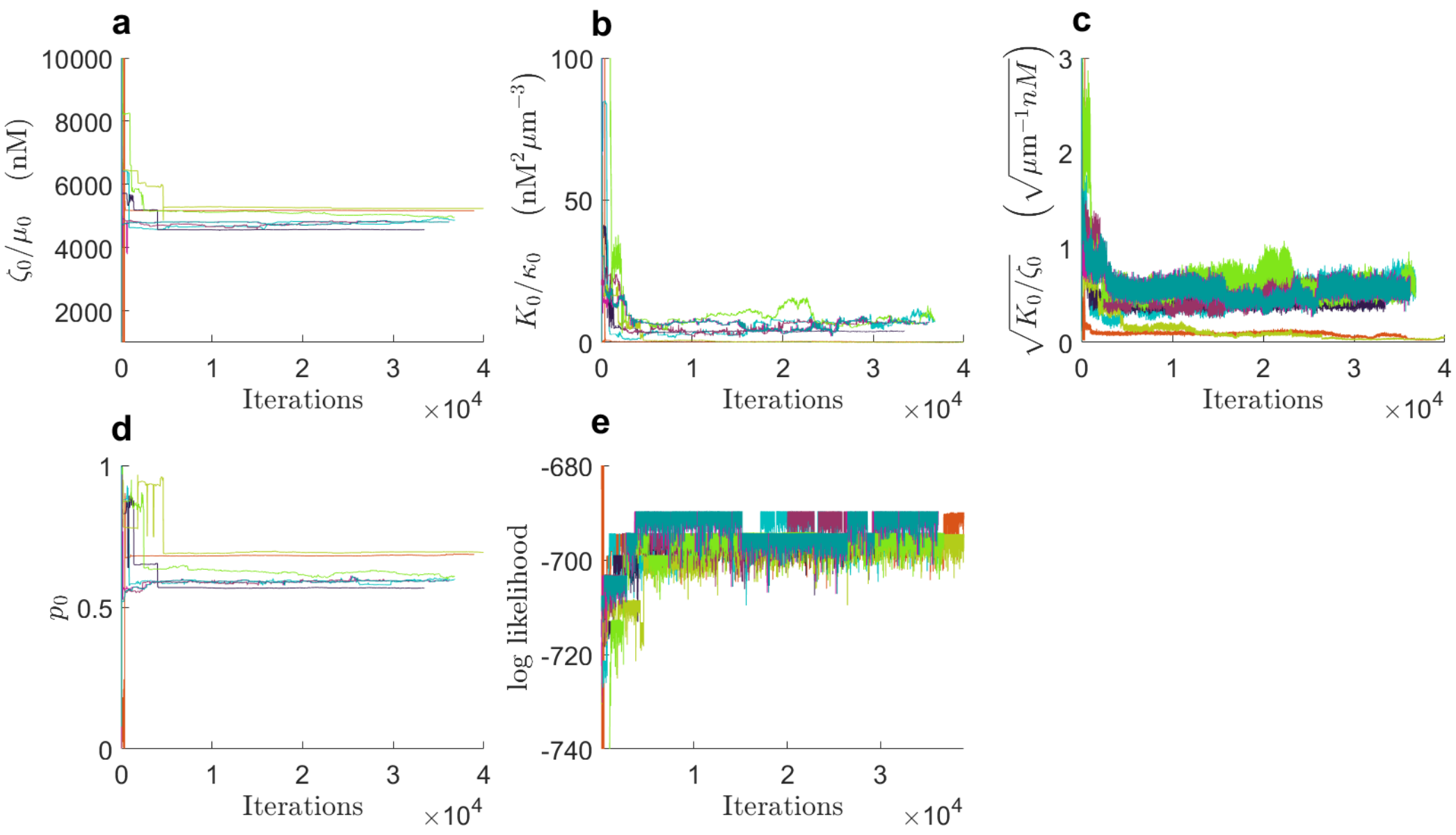

Fig. S11: **Parameter estimation through Markov Chain Monte Carlo**. Time series for the updated values of the fit parameters a) r1, b) r2, c) r3, and d) p0. We started 10 simulations from the upper end of the estimated parameter range, and another 10 from the lower end of the range. All the runs attained similar parameter values on convergence e) likelihood associated with the parameters displayed in a-d. Each colored curve corresponds to a different simulation. The parameter values we report are associated with the run that attained the largest log likelihood value on convergence (teal colored)

# 6 Supplementary Video Captions

**Video S1: In-plane bend instability of an aligned suspension of microtubules, crosslinkers and molecular motor clusters.**

Widefield fluorescent microscope, 10X objective, [ATP]=50$\mu$M, [motor clusters]=20$nM$, [PRC1]=100$nM$.

**Video S2: In-plane bend instability of an aligned suspension of microtubules, depletant, and molecular motor clusters.**

Widefield fluorescent microscope, 4X magnification, [ATP]=1$mM$, [motor clusters]=20$nM$, [PEG 20kDa]=0.8% (vol/vol).

**Video S3: Out-of-plane buckling of an aligned suspension of microtubules, crosslinkers and molecular motor clusters.**

Widefield fluorescent microscope, [ATP]=10$\mu$M, [motor clusters]=10nM, [PRC1]=200$nM$.

**Video S4: Out-of-plane buckling of an aligned suspension of microtubules, depletant, and molecular motor clusters.**

Widefield fluorescent microscope, 10X magnification, [ATP]=10$\mu$M, [motor clusters]=20nM, [[PEG 20kDa]=0.8% (vol/vol).

**Video S5: Confocal time-lapse imaging of the in-plane bend instability.**

[ATP]=130$\mu$M, [motor cluster]=15$nM$, [PRC1]=50$nM$. Microtubule Length=1.5$\mu$m, Zstep 2.5$\mu$m, 1min23sec deltaT 20x

**Video S6: Confocal time-lapse imaging of the out-of-plane buckling.**

[ATP]=4$\mu$M, [motor cluster]=3$nM$, [PRC1]=100$nM$, Microtubule Length=2.3$\mu$m, Zstep size: 2$\mu$m, delatT 4min, 20X.

**Video S7: Segmentation of the out-of-focus regions of interest.**

First row: in-plane instability, second row: superposition of in-plan and out-of-plane deformations, third raw: out-of-plane buckling. The red areas correspond to the parts of the network that are out-of-focus.

**Video S8: turning off motor activity only allow a partial relaxation of the deformations.**

Fluorescent timelapse imaging of the microtubule bundles 30sec after blue light is turned off.

# References

[1] P. G. d. Gennes and J. Prost, *The physics of liquid crystals.* Birman, J., series joint gen. ed. International series of monographs on physics, Oxford : New York: Clarendon Press ; Oxford University Press, 2nd ed. ed., 1998.

[2] R. Aditi Simha and S. Ramaswamy, "Hydrodynamic fluctuations and instabilities in ordered suspensions of self-propelled particles," *Phys. Rev. Lett.*, vol. 89, p. 058101, Jul 2002.

[3] T. Sanchez, D. T. N. Chen, S. J. DeCamp, M. Heymann, and Z. Dogic, "Spontaneous motion in hierarchically assembled active matter," *Nature*, vol. 491, no. 7424, pp. 431–434, 2012.

[4] S. P. Gilbert, M. R. Webb, M. Brune, and K. A. Johnson, "Pathway of processive atp hydrolysis by kinesin," *Nature*, vol. 373, pp. 671–676, Feb 1995.

[5] S. P. Gilbert, M. L. Moyer, and K. A. Johnson, "Alternating site mechanism of the kinesin atpase," *Biochemistry*, vol. 37, pp. 792–799, Jan 1998.

[6] S. P. Gilbert and K. A. Johnson, "Pre-steady-state kinetics of the microtubule-kinesin ATPase," *Biochemistry*, vol. 33, pp. 1951–1960, Feb. 1994.

[7] P. Xie, "Theoretical analysis of dynamics of kinesin molecular motors," *ACS Omega*, vol. 5, pp. 5721–5730, Mar 2020.

[8] M. L. Gardel, J. H. Shin, F. C. MacKintosh, L. Mahadevan, P. Matsudaira, and D. A. Weitz, "Elastic behavior of cross-linked and bundled actin networks," *Science*, vol. 304, no. 5675, pp. 1301–1305, 2004.

[9] Y. Luan, O. Lieleg, B. Wagner, and A. R. Bausch, "Micro- and macrorheological properties of isotropically cross-linked actin networks," *Biophysical Journal*, vol. 94, no. 2, pp. 688–693, 2008.

[10] O. Lieleg, M. M. A. E. Claessens, C. Heussinger, E. Frey, and A. R. Bausch, "Mechanics of bundled semiflexible polymer networks," *Phys. Rev. Lett.*, vol. 99, p. 088102, Aug 2007.

[11] B. Wagner, R. Tharmann, I. Haase, M. Fischer, and A. R. Bausch, "Cytoskeletal polymer networks: The molecular structure of cross-linkers determines macroscopic properties," *Proceedings of the National Academy of Sciences*, vol. 103, no. 38, pp. 13974–13978, 2006.

[12] Y.-C. Lin, G. H. Koenderink, F. C. MacKintosh, and D. A. Weitz, "Viscoelastic properties of microtubule networks," *Macromolecules*, vol. 40, no. 21, pp. 7714–7720, 2007.

[13] J. H. Shin, M. L. Gardel, L. Mahadevan, P. Matsudaira, and D. A. Weitz, "Relating microstructure to rheology of a bundled and cross-linked f-actin network in vitro," *Proceedings of the National Academy of Sciences*, vol. 101, no. 26, pp. 9636–9641, 2004.

[14] J. P. Straley, "Critical phenomena in resistor networks," *Journal of Physics C: Solid State Physics*, vol. 9, pp. 783–795, mar 1976.

[15] E. E. Magat, "Liquid crystallinity in polymers, principles and fundamental properties, alberto ciferri, ed., vch, new york, 1991, 438 pp.," *Journal of Polymer Science Part A: Polymer Chemistry*, vol. 30, no. 5, pp. 955–955, 1992.

[16] P. W. Ellis, D. J. G. Pearce, Y.-W. Chang, G. Goldsztein, L. Giomi, and A. Fernandez-Nieves, "Curvature-induced defect unbinding and dynamics in active nematic toroids," *Nature Physics*, vol. 14, pp. 85–90, Jan 2018.

[17] D. A. Gagnon, C. Dessi, J. P. Berezney, R. Boros, D. T.-N. Chen, Z. Dogic, and D. L. Blair, "Shear-induced gelation of self-yielding active networks," *Phys. Rev. Lett.*, vol. 125, p. 178003, Oct 2020.

[18] M. Castoldi and A. V. Popov, "Purification of brain tubulin through two cycles of polymerization–depolymerization in a high-molarity buffer," *Protein Expression and Purification*, vol. 32, pp. 83–88, Nov 2003.

[19] G. Duclos, R. Adkins, D. Banerjee, M. S. Peterson, M. Varghese, I. Kolvin, A. Baskaran, R. A. Pelcovits, T. R. Powers, A. Baskaran, *et al.*, "Topological structure and dynamics of three-dimensional active nematics," *Science*, vol. 367, no. 6482, pp. 1120–1124, 2020.

[20] S. J. DeCamp, G. S. Redner, A. Baskaran, M. F. Hagan, and Z. Dogic, "Orientational order of motile defects in active nematics," *Nature materials*, vol. 14, no. 11, pp. 1110–1115, 2015.

[21] R. Subramanian, E. M. Wilson-Kubalek, C. P. Arthur, M. J. Bick, E. A. Campbell, S. A. Darst, R. A. Milligan, and T. M. Kapoor, "Insights into antiparallel microtubule crosslinking by prc1, a conserved nonmotor microtubule binding protein," *Cell*, vol. 142, no. 3, pp. 433–443, 2010.

[22] D. S. Martin, R. Fathi, T. J. Mitchison, and J. Gelles, "Fret measurements of kinesin neck orientation reveal a structural basis for processivity and asymmetry," *Proceedings of the National Academy of Sciences*, vol. 107, no. 12, pp. 5453–5458, 2010.

[23] T. D. Ross, H. J. Lee, Z. Qu, R. A. Banks, R. Phillips, and M. Thomson, "Controlling organization and forces in active matter through optically defined boundaries," *Nature*, vol. 572, no. 7768, pp. 224–229, 2019.

[24] P. Chandrakar, M. Varghese, S. Aghvami, A. Baskaran, Z. Dogic, and G. Duclos, "Confinement controls the bend instability of three-dimensional active liquid crystals," *Phys. Rev. Lett.*, vol. 125, p. 257801, Dec 2020.

[25] A. Desai, S. Verma, T. J. Mitchison, and C. E. Walczak, "Kin i kinesins are microtubule-destabilizing enzymes," *Cell*, vol. 96, no. 1, pp. 69–78, 1999.

[26] R. Williams and L. A. Rone, "End-to-end joining of taxol-stabilized gdp-containing microtubules," *Journal of Biological Chemistry*, vol. 264, no. 3, pp. 1663–1670, 1989.

[27] S. A. Aghvami, A. Opathalage, Z. Zhang, M. Ludwig, M. Heymann, M. Norton, N. Wilkins, and S. Fraden, "Rapid prototyping of cyclic olefin copolymer (coc) microfluidic devices," *Sensors and Actuators B: Chemical*, vol. 247, pp. 940–949, 2017.

[28] T. Sanchez, D. T. Chen, S. J. DeCamp, M. Heymann, and Z. Dogic, "Spontaneous motion in hierarchically assembled active matter," *Nature*, vol. 491, no. 7424, pp. 431–434, 2012.

[29] A. D. Edelstein, M. A. Tsuchida, N. Amodaj, H. Pinkard, R. D. Vale, and N. Stuurman, "Advanced methods of microscope control using Œºmanager software," *Journal of Biological Methods*, vol. 1, p. e10, Nov. 2014.

[30] W. Thielicke and E. J. Stamhuis, "Pivlab-time-resolved digital particle image velocimetry tool for matlab," *Published under the BSD license, programmed with MATLAB*, vol. 7, no. 0.246, p. R14, 2014.